\documentclass{emulateapj}

\usepackage{amsmath}
\usepackage{amssymb}
\usepackage{epsfig}
\usepackage{graphicx}
\usepackage{ifthen}
\usepackage{latexsym}
\usepackage{rotating}
\usepackage{subfigure}
\usepackage{times,epsf}
\usepackage{txfonts}
\usepackage{varioref}
\usepackage{verbatim}
\usepackage{url}
\usepackage{color}
\usepackage[dvipsnames]{xcolor}
\usepackage[T1]{fontenc}
\usepackage{float}
\usepackage[normalem]{ulem}

\usepackage{bm}

\begin{document}

\begin{abstract}
The reprocessing of primary X-ray emission in the accretion disk of black hole X-ray binaries (BHXRBs) produces a reflection spectrum with the characteristic Fe K$\alpha$ fluorescence line. Strong low-frequency quasi-periodic oscillations (QPOs) are observed from BHXRBs, and the dependence of QPO properties (e.g., phase lag) on the inclination angle suggests that the observed QPO may be associated with a geometrical effect, e.g., the precession of the X-ray source due to frame dragging near the spinning black hole. Here, in the scenario of the Lense-Thirring precession of the X-ray source, we use a Monte Carlo simulation of radiative transfer to study the irradiation/reflection and the resultant spectral properties including the Fe K$\alpha$ line as a function of precession phase (time). We found that the reflection fraction, i.e., the ratio of the incident flux toward the disk and the direct flux toward to the observer at infinity, is modulated by the precession phase, which depends on the truncation radius (i.e., the spectral state in the truncated disk model) and the inclination angle. The Fe K$\alpha$ line profile also changes as the primary X-ray source precesses, with the line luminosity and the flux-weighted centroid energy varying with the precession phase. The periodically modulated 2-10 keV continuum flux could apparently lag the line luminosity in phase, if the truncation radius is small enough for Doppler effects due to disk orbital motion to significantly affect the observed radiation.

\end{abstract}


%\title{X-ray quasi-periodic oscillations in Lense--Thirring precession model-I.  relativistic continuum variability}
\title{X-ray quasi-periodic oscillations in the Lense--Thirring precession model - II.  variability of the relativistic iron K$\alpha$ Line}

\author{Bei You\altaffilmark{1,2}, Piotr T. \.{Z}ycki\altaffilmark{3}, Adam Ingram\altaffilmark{4}, Michal Bursa\altaffilmark{5}, Wei Wang\altaffilmark{1,2}}

\altaffiltext{1}{School of Physics and Technology, Wuhan University, Wuhan 430072, China; youbei@whu.edu.cn}
\altaffiltext{2}{Astronomical Center, Wuhan University, Wuhan 430072, People’s Republic of China}
\altaffiltext{3}{Nicolaus Copernicus Astronomical Center, Polish Academy of Sciences, Bartycka 18, 00-716 Warsaw, Poland}
\altaffiltext{4}{Department of Physics, Astrophysics, University of Oxford, Denys Wilkinson Building, Keble Road, Oxford, OX1 3RH, UK}
\altaffiltext{5}{Astronomical Institute, Academy of Sciences, Bo\v cn\'i II 1401, 14131 Prague, Czech Republic}

\maketitle

\section{Introduction}

%In the hard state of black hole binary, the accretion thin disk is thought to be truncated at a radius larger than innermost stable circular orbit (ISCO), while below the truncation radius it is the hot flow. Due to the frame-dragging effect of the spinning BH, the inner flow will undergo the Lense-Thirring precession if the angular momentum of the flow misalign with BH spin. This Lense-Thirring precession is one of the interpretations of the low frequency QPO. 

In black hole X-ray binaries (BHXRBs), soft seed photons from the accretion disk are inverse Comptonized by hot electrons near the black hole (BH), contributing to the observed Comptonization component in the spectrum at high energy (Galeev et al.1979; Haardt \& Maraschi 1991; Cao 2009; Qiao \& Liu 2012; You et al. 2012). A fraction of the Comptonized photons will irradiate the accretion disk where they are absorbed and/or scattered to produce the reflection emission. The characteristic features of the reflection emission are the hump between $\sim$ 20-30 keV and the Fe K$\alpha$ emission line at around $E=6.4$ keV in the rest frame (George \& Fabian 1991; You et al. 2018; Zhang et al. 2019).

One simple reflection geometry is the lamp-post model,
%in which the illumination source in the power law $F_{\rm E} \sim E^{\alpha}$, 
in which the accretion disk extends down to some radius close to the BH, whereas the illuminating X-ray source is assumed to be a point source located on the spin axis of the BH, at a certain height (Miniutti \& Fabian 2004; Nied{\'z}wiecki et al. 2016; Vincent et al. 2016). This X-ray illumination source could also be vertically extended (Kara et al. 2019). 
Another possible reflection geometry is the truncated disk model, in which the accretion disk is truncated at some radius $R_{\rm tr}$ which could be larger than the innermost stable circular orbit (ISCO). Interior to $R_{\rm tr}$ is the spatially extended illuminating source, which is of high temperature, acting as the Comptonized component (Done, Gierli{\'n}ski, \& Kubota 2007; Plant et al. 2015; Basak \& Zdziarski 2016; Mahmoud et al. 2019).
This truncated disk geometry has been mentioned in the literature to explain the spectral transition and the evolution of the disk reflection properties (Esin et al. 1997; Gilfanov 2010). Recently, this truncation geometry coupled with the so-called Lense-Thirring precession was considered as the possible origin of the observed
quasi-periodic oscillation (QPO) in BHXRBs (Stella \& Vietri 1998; Motta et al. 2015; Stiele \& Yu(2016); van den Eijnden et al. 2017; Huang et al.2018; Xiao et al. 2019; Zhang et al. 2020; see Belloni \& Motta 2016, and Ingram \& Motta 2019, for reviews of alternative QPO models).
It is the inner hot flow (hereafter the corona) that precesses around the BH axis due to the frame-dragging effect, periodically modulating the observed X-ray flux (Ingram et al. 2009; You et al. 2018). 
%Therefore, from the perspective of fast variability (i.e., QPO), the corona as the Comptonization source near BH is required to not only have the spcially extended distribution (rather point source) but also be in the dynamical motion (i.e., the precession in the azimuthal direction). 
In this case, the illumination pattern of the precessing flow on the disk will also vary as a function of time. This will eventually result in the modulation of the reflection including the Fe K$\alpha$ line. Ingram \& Done (2012) investigated the effect of precession of the illumination pattern on the resultant Fe K$\alpha$ line. Given the rotation of the accretion disk, when the approaching side of the disk is predominantly illuminated by the X-ray photons from the flow, the resultant iron line will be blueshifted and boosted; when the receding side of the disk is illuminated, the iron line will be predominantly redshifted. As the illumination pattern on the disk rotates due to the precession of the flow, the overall shape and peak flux of of the iron line from the entire disk will periodically rock between blueshift and redshift.
Observationally, the Fe K$\alpha$ line has been reported to be variable in some BHXRBs, e.g., GRS 1915+105 (Miller \& Homan 2005). Ingram et al. (2016, 2017, hereafter ID17) applied a phase-resolved analysis on XMM-Newton and NuSTAR observations of H1743-322 and found that both the Fe K$\alpha$ line centroid energy and the reflection fraction vary systematically as a function of QPO phase, with high significance, which suggesting that the observed QPO is produced by precession.

The phase-dependent behavior of the Fe K$\alpha$ line in the Lense-Thirring precession model is a powerful diagnostic of not only the geometry but also the dynamics of the accretion flow very close to the BH. This means that the analysis of the reflection component including the Fe K$\alpha$ line as a function of QPO phase during the outburst can be used to study the evolution of the accretion flow geometry and relativistic effect in BHXRBs.
However, this potential cannot be realized without quantitatively 
%prerequisite would be realistic 
modeling the phase-dependent reflection component including the Fe K$\alpha$ line in the Lense-Thirring precession model.
For the scenario of the truncated disk, You et al. (2018) developed a Monte Carlo code to deal with the radiative transfer process for photons in the precessing hot flow to simulate the QPO variability arising from the Lense-Thirring precession in the framework of full general relativity. In our simulation, the code is able to self-consistently simulate both Comptonization and reflection (including Fe K$\alpha$) processes. 
The contribution of the spectral components, i.e., the disk emission, the Comptonization, and the reflection emission, to the variability was studied, which allows us to study the overall QPO variability as a function of photon energy. More importantly, by using the Monte Carlo code, You et al. (2018) quantitatively studied the  evolution of the QPO variability during the spectral transition from the hard to soft states, and the correlation between the QPO variability and QPO frequency. 
%The QPO properties of the continuum emission, mainly concentrating on the fractional variability amplitude, was stuied in details in You et al. (2018).
In this work, we study the QPO phase-dependent properties of disk reflection, including the Fe K$\alpha$ line, as a function of time (precession phase), in the Lense-Thirring precession.
%It is noteworthy that the pioneer work was done by Ingram et al. (2012) to analytically investigate the QPO behaviour of Fe Ka line in the truncated disk + Lense-Thirring precession model, under a few assumptions. The improvment of our simulation code in You et al. (2018) based on their work is that we adopt the Monte-Carlo strategy to treat the radiative transfer of each single photon in the accretion flow, namely the reflection within the outer code disk and the Comptonization in the corona. Meanwhile, since the relative process is near BH where GR is inignore, we take the light-bending and gravitational redshift into account.

A general description of our simulation will be summarized in Sect. 2. 
%Modelling of the QPO of the Fe Ka line for the fidual parameters, 
The modeling of the illumination/reflection patterns on the disk and the resultant Fe K$\alpha$ line, as a function of the precession phase, will be shown in Sect. 3.
%More modellng results will be given in Sect.4 for different cases of the model parameters, 
%in order to directly analysize the effect of both truncation radius, general relativity and observer position () on the variation of Fe Ka line.
The phase lag between the spectral properties and the effect of the observer position at infinity on the QPO properties will be discussed in Sect. 4. The main conclusions of this work are listed in Sect. 5.

\section{model}

\subsection{Geometry configuration}

\begin{figure}
\includegraphics[width=\columnwidth]{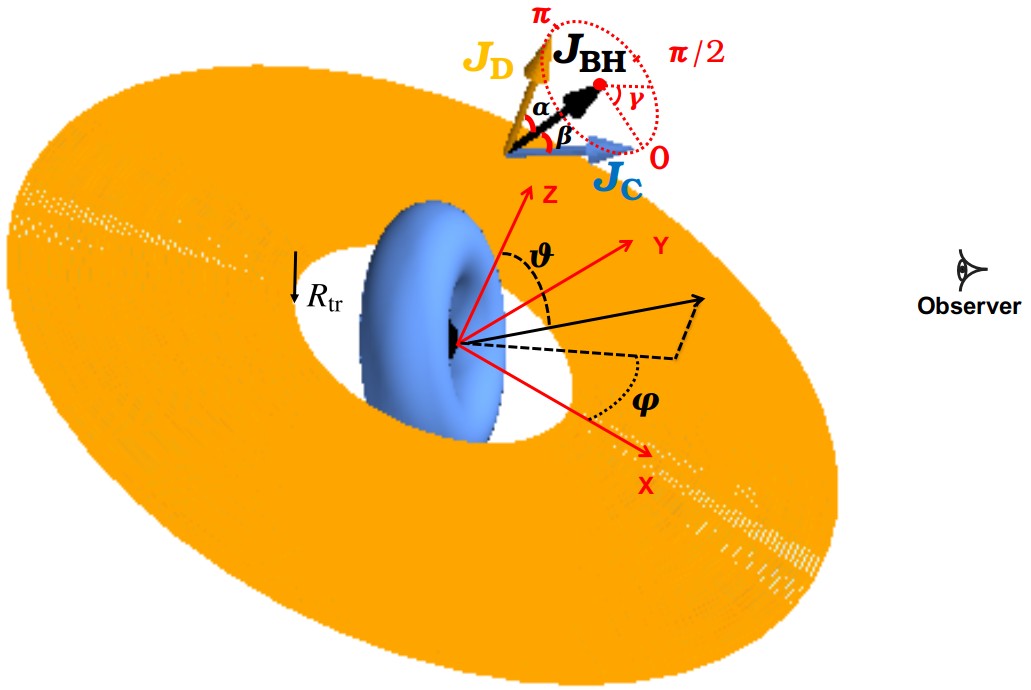}
\caption{
Schematic of the assumed geometry of the accretion flow and coordinates in this paper. The outer thin accretion disk (the plane in yellow) is truncated at $\rm R_{\rm tr}$ which is also assumed to be the outer radius of the precessing corona (torus-like shape in blue). The axis vector of the outer disk $\bm{J_{\rm D}}$ is misaligned with that of the black hole $\bm{J_{\rm BH}}$ by an angle  $\alpha$. We define {\bf the beginning phase} as the moment when the axis vector of the corona $\bm{J_{\rm C}}$ is on the plane confined by ($\bm{J_{\rm BH}}$, $\bm{J_{\rm D}}$) and the 
the axis vector of the corona $\bm{J_{\rm C}}$ is misaligned with $\bm{J_{\rm BH}}$ by an angle $\beta$. It is simply assumed that $\alpha = \beta$. As the corona undergoes Lense-Thirring precession, the vector $\bm{J_{\rm C}}$ circularly rotates around the vector $\bm{J_{\rm BH}}$, with the angle $\gamma$ with respect to the beginning phase, varying from 0 to $2\pi$.
The Cartesian coordinate ($X$, $Y$, $Z$) is defined in such a way that the Z-axis is directed toward $\bm{J_{\rm D}}$, and the X-axis is on the plane confined by ($\bm{J_{\rm BH}}$, $\bm{J_{\rm D}}$), pointing to the misalignment direction. The viewing angle of the observer $\theta$ is defined with respect to the Z-axis, ranging from 0 and $\pi /2$, and the azimuth of the observer $\varphi$ is defined with respect to the X-axis. The BH spin $a=0.3$ is assumed.
\label{schematic}}
\end{figure}

In this work, we use a Monte Carlo simulation of radiative transfer to 
simulate the phase-resolved emission, e.g., the Comptonization and the reflection including the characteristic Fe K$\alpha$ line, in the Lense-Thirring precession model for the BHXRB. 
Fig. \ref{schematic} shows the assumed geometry of the accretion flow. The outer thin accretion disk is truncated at $R_{\rm tr}$ which is also assumed to be the outer radius of the precessing corona. The inner radius of the corona is $R_{\rm in}$. The axis vector of the outer disk $\bm{J_{\rm D}}$ is misaligned with that of the BH $\bm{J_{\rm BH}}$ by an angle $\alpha$. We define {\bf the beginning phase} as the moment when the axis vector of the corona $\bm{J_{\rm C}}$ is on the plane confined by ($\bm{J_{\rm BH}}$, $\bm{J_{\rm D}}$) and the axis vector of the corona $\bm{J_{\rm C}}$ is misaligned with $\bm{J_{\rm BH}}$ by an angle $\beta$, as is pictured in Fig. \ref{schematic}. As the corona undergoes Lense-Thirring precession, the vector $\bm{J_{\rm C}}$ circularly rotates around the vector $\bm{J_{\rm BH}}$, 
%with the angle $\gamma$ with respect to the beginning phase, 
with phase given by the precession angle $\gamma$, ranging from 0 to $2\pi$. 
The maximum misalignment happens when the corona is at the beginning phase of the precession with $\gamma = 0$. For simplicity, we assume $\alpha = \beta$, so that the corona aligns with the disk at the middle phase when $\gamma = \pi$. 

Moreover, we define the Cartesian coordinates ($X$, $Y$, $Z$) in such a way that the Z-axis is directed along $\bm{J_{\rm D}}$, and the X-axis is on the plane confined by ($\bm{J_{\rm BH}}$, $\bm{J_{\rm D}}$), pointing to the misalignment direction. The viewing angle of the observer $\theta$ is defined with respect to the Z-axis, ranging from 0 and $\pi /2$, and the azimuth of the observer $\varphi$ is defined with respect to the X-axis. 
%In the simulation of this work, we simulate the observed phase-resolved emission when the observer azimuth $\varphi = 0, \pi /2, \pi, 3\pi /2$. 
At the beginning precession phase $\gamma/2\pi = 0$, when $\varphi = 0$ or $\pi$, we observe the largest projected area of the corona. {\bf Note that the simulation results in the Sect. 3 are for $\varphi = 0$}. The effect of the observer azimuth on the phase-resolved emission will be discussed in Sect. 4. 

Based on this geometry, we will simulate the phase-resolved spectrum from the system as a function of precession phase $\gamma/2\pi$, by using our relativistic radiative transfer code (You et al. 2018). The strategy of the simulation code is summarized as follows: 
\begin{itemize}
\item First, randomly initializing seed photons from the outer disk, which are required to obey the blackbody emission from the relativistic accretion disk (Novikov \& Thorne 1973);
\item Second, determining the destination of each individual seed photon:
  \begin{itemize}
  \item (i) if going directly to infinity, then being saved as disk spectrum;
  \item (ii) if going to the corona, then being inverse-Compton scattered. The process of inverse Comptonization is implemented by following the prescriptions of Pozdnyakov et al. (1983) and Gorecki \& Wilczewski (1984; see also Janiuk et al. 2000). If the Comptonized photons escaping from the corona travel to infinity, then it will be saved as the Comptonization spectrum (as the {\bf direct component}).
  \end{itemize}
\item For the Comptonized photons escaping from the corona, if they illuminate the disk (as the {\bf incident component}), they will be reprocessed and reflected back to the infinity, being saved as the contribution to  the reflection spectrum. The reprocessing of the Comptonized photon in the disk is implemented by following the prescriptions of George \& Fabian (1991) and Zycki \& Czerny (1994). In this case, the predominant fluorescent photon at 6.4 keV, i.e., the Fe K$\alpha$ line, is simulated as well.   
\end{itemize}

More details on the simulation process can be found in You et al.(2018) in which the variability of the continuum due to the precession of the inner hot flow was studied.
In this work, we will focus on the phase-resolved emission from the system, mainly concentrating on the reflection and the Fe K$\alpha$ emission line, in the Lense-Thirring precession model.

\subsection{Spectral state transitions}

The electron temperature and the optical depth are the key parameters in shaping the Comptonization spectrum in terms of the spectral slope and the cutoff at high energy (Zdziarski et al. 1996; Qiao \& Liu 2018; Yan et al. 2020). 
In our simulation, in order to implement the spectral transition, in the scenario of truncated disk geometry, we use the \textit{eqpair} code\footnote{http://www.astro.yale.edu/coppi/eqpair/} to calculate the electron temperature and the optical depth for different truncation radii (You et al., 2018).

Given a truncation radius, the ratio of the hardness and softness compactness $l_{\rm h}/l_{\rm s}$ (where $l_{\rm h,\rm s} = L_{\rm h,\rm s}\sigma_{\rm T}/Rm_{\rm e}c^3$, and $L_{\rm h,\rm s}$ is the source luminosity) and the value of $l_s$ are the key inputs in \textit{eqpair}, as these two parameters dominantly determine the electron energy distribution, which is solved by the balance between the heating (including direct acceleration of particles) and cooling processes, including electron-positron pair balance,
bremsstrahlung and Compton cooling (Poutanen \& Coppi 1998; Coppi 1999).
We refer the reader to You et al., (2018) for the detailed computations of $l_{\rm h}/l_{\rm s}$ and $l_s$.

Under the geometric configuration of the disk/corona in Sect. 2.1, both $l_{\rm h}/l_{\rm s}$ and $l_s$ can be estimated, which have monotonous relationships with the truncation radius. Then, the parameters of the Comptonizing plasma (optical depth $\tau$, phase-averaged electron temperature $T_{\rm e}$) can be computed for a given $R_{\rm tr}$. More specifically, when the truncation radius $R_{\rm tr} = 90$, then $T_{\rm e} \simeq 110$ keV, $\tau \simeq 2.1$, and $\Gamma \simeq 1.3$; when the truncation radius $R_{\rm tr} = 10$, then $T_{\rm e} \simeq 85$ keV, $\tau \simeq 1.0$, and $\Gamma \simeq 1.7$. It was suggested in previous studies that the density and the optical depth might be radially stratified (Axelsson et al. 2013; Axelsson et al. 2014), which may affect the QPO properties. Studying the effect of the inhomogeneous corona on the QPO properties is beyond the scope of this work. Therefore, in this work, we simply assume that the corona is homogeneous with the constant density and take the optical depths above as those of the corona which are radially integrated from the inner radius $R_{\rm in}$ to the outer radius $R_{\rm tr}$. 
The optical depth $\tau$ and electron temperature $T_{\rm e}$ are taken as input parameters in our Monte Carlo simulation to compute the X-ray energy spectrum and the associated variability in the following sections.

\section{results}
In the truncated disk geometry, which is assumed in this work, seed photons from the outer disk are Comptonized in the corona. The Comptonized photons can escape out of the corona, some of which irradiate the outer disk, producing the reflection spectrum and the characteristic Fe K$\alpha$ line as well. 
Therefore, in the scenario of Lense-Thirring precession, in order to study the variation of the disk reflection and Fe K$\alpha$ line with the precession phase (Sect. 3.2 and 3.3), it is essential to first study the variation of the irradiation (Sect. 3.1). 

\subsection{Variation of the irradiation}

\subsubsection{The case of the Lamp post}
\begin{figure}
\includegraphics[width=\columnwidth]{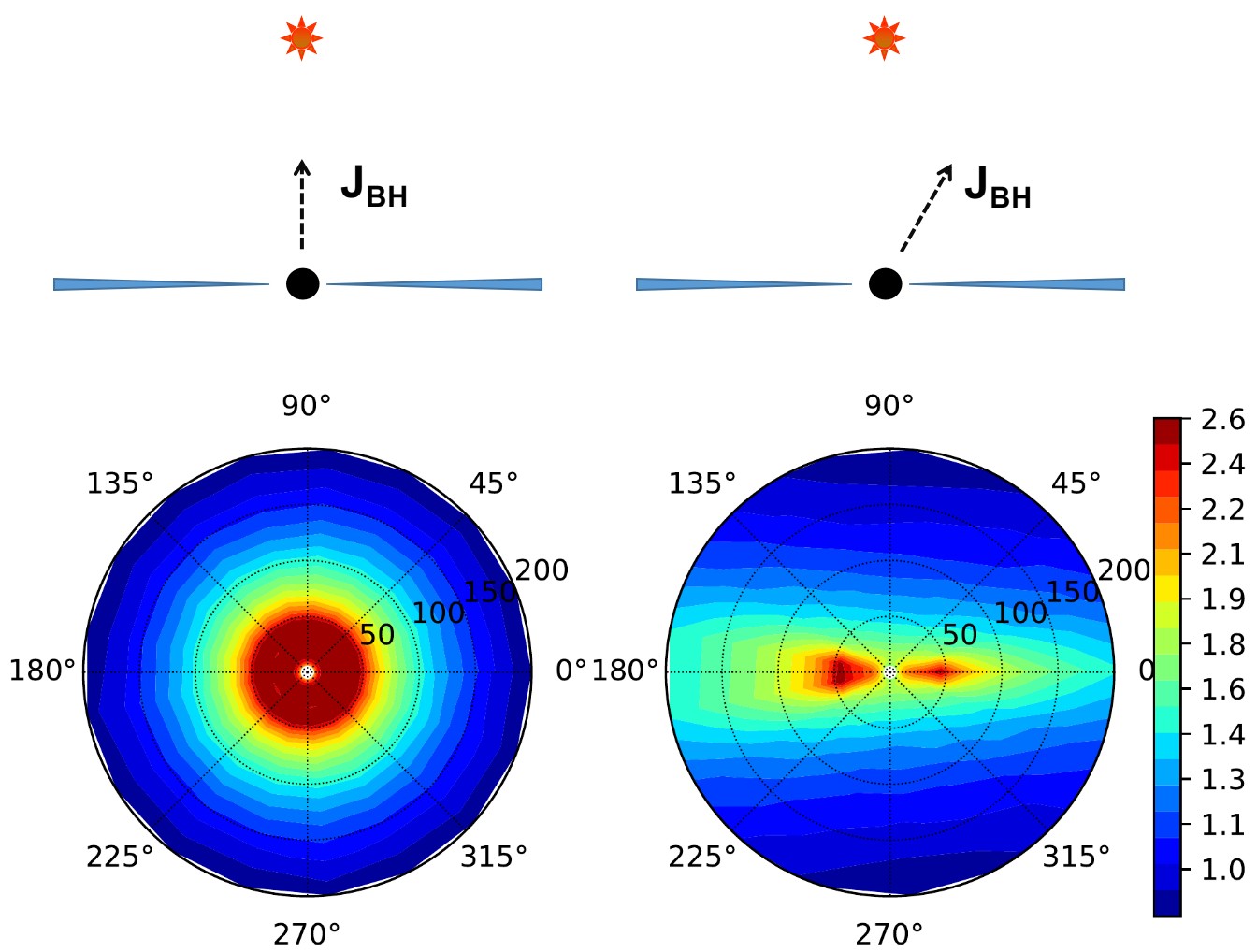}
\caption{
Irradiation pattern (in the disk rest frame) for the lamp-post geometry in which the X-ray source (corona) is assumed to be a point-like source above the disk. Upper left: the BH axis is aligned with the disk axis. Bottom left: the resultant irradiation pattern. Upper right: the BH axis is misaligned with the disk axis. Bottom right: the resultant irradiation pattern. The color bar represents the scaled photon counts (in logarithm) intercepted by the disk. It can be seen that the difference in intercepted photon counts could be two orders of magnitude. Note that the results are in the disk rest frame. The BH spin $a=0.3$ is assumed, and the inner radius of the accretion disk is assumed to be at the innermost stable circular orbit, i.e., $R_{\rm tr} = R_{\rm ISCO}$. 
\label{lamp_post}}
\end{figure}

In order to study the effect of the misalignment between the BH spin axis and the illuminating corona on the irradiation pattern over the disk, we first explore the simple case of the corona, i.e., the lamp-post geometry. In this case, the corona, as the X-ray source, is assumed to be a point-like illuminating source which is located on {\bf the disk symmetry axis} above the disk at a height of $H$ (see the schematic in Fig. \ref{lamp_post}). The illuminating photons from the corona are intercepted by the disk. When the BH spin is aligned with the disk axis, the irradiation pattern observed in the disk rest frame, i.e., the distribution of the illuminating photons over the disk, is perfectly symmetric, as expected. However, when the BH spin is misaligned with the disk axis, tilting toward the X-axis, as shown in the right panel of Fig. \ref{lamp_post}, the axisymmetry of the irradiation pattern turns out to be broken due to the lens effect (light-bending effect) on the photon trajectory, which is prolonged along the X-axis. The color bar represents the scaled photon counts (in logarithm) intercepted by the disk. It can be seen that the difference in the counts of the intercepted photons could be two orders of magnitude.
More importantly, we find that two bright patches are formed on the X-axis, with the left side being brighter than the right one. If this point-like X-ray source (corona) is vertically extended, e.g., a jet-like base (Wilkins et al. 2015; Kara et al. 2019; You et al. 2020, in preparation), and precesses around the BH spin axis with some period (Liska et al. 2018), then we would expect these two bright patches on the disk to vary, which would eventually result in the periodic modulation of the reflection flux at about half of the precession period. This effect may be responsible for the observed harmonic in the power spectrum of BHXRBs (Axelsson et al. 2014; Axelsson \& Done 2016; Stevens \& Uttley 2016; Huang et al.2018; de Ruiter et al. 2019; Xiao et al. 2019; Stevens et al. 2018), which will be further studied in a subsequent paper.  

\subsubsection{The case of the precessing corona}
Now we consider another geometry of the X-ray irradiating source, i.e., the extended torus-like corona, while the disk is truncated at some radius, which is of our research interest in this work. 
The corona is assumed to undergo Lense-Thirring precession (counterclockwise) around the BH spin axis (see Fig. \ref{schematic}), and the outer disk is assumed to rotate counterclockwise as well.
In this case, the irradiation pattern is found to be different from that of the lamp-post case and shows the expected variation as the corona precesses.

\begin{figure}
\includegraphics[width=\columnwidth]{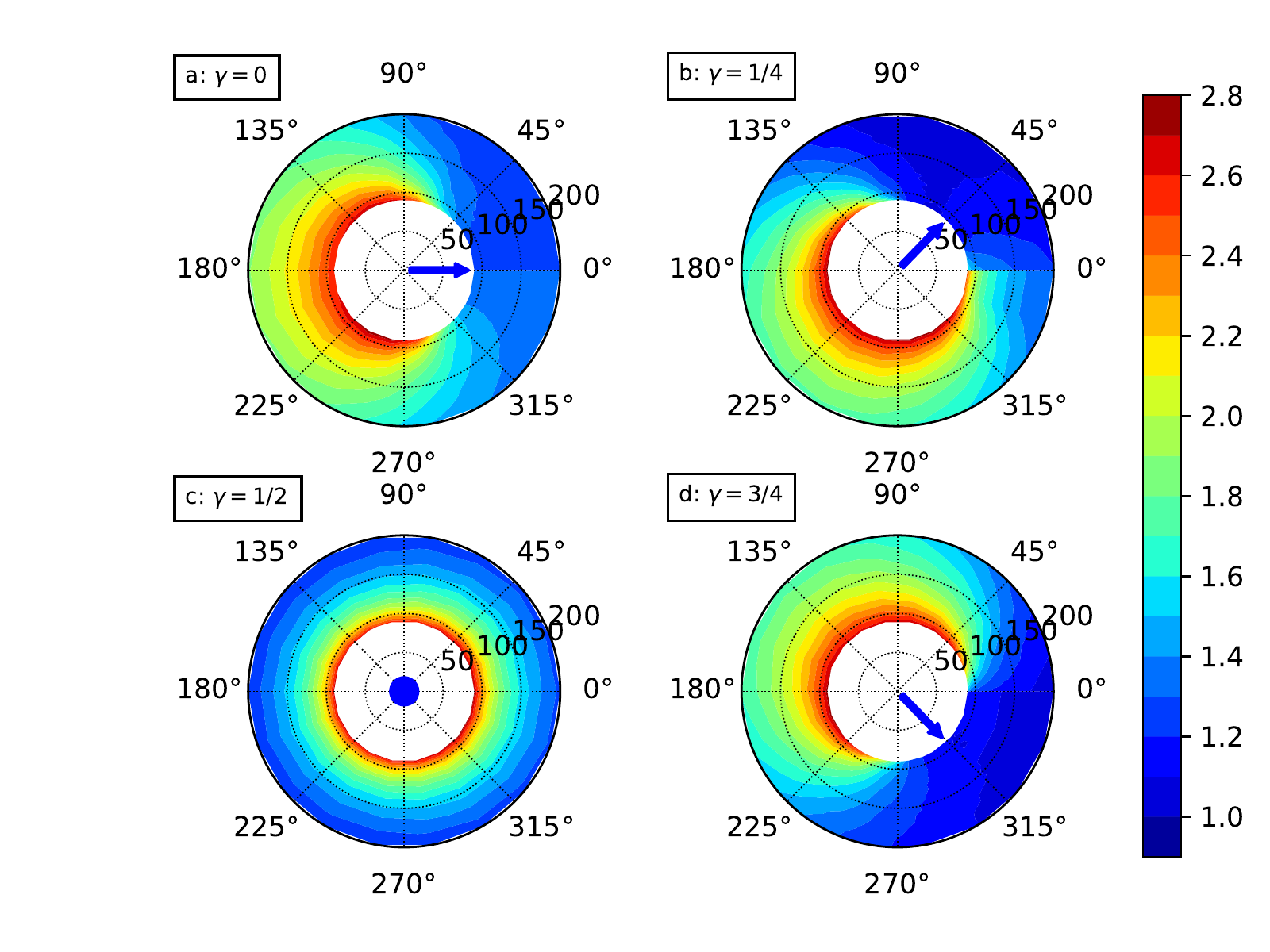}
\caption{
Incident pattern in the disk restframe, in terms of luminosity $L = \int F_E {\rm d}E$, for the case of a truncated disk plus precessing corona (i.e., Fig. \ref{schematic}). The truncation radius $R_{\rm tr} = 90$. The BH spin $a=0.3$ is assumed. The color bar represents the scaled photon counts (in logarithm) intercepted by the disk. The difference in intercepted luminosity could be up to 1.8 orders of magnitude. The four panels (a), (b), (c), and (d) correspond to the four precession angles $\gamma/2\pi = 0$, $1/4$, $1/2$, and $3/4$, respectively. In each panel, the blue arrow at the center represents the projection of the corona axis $\bm{J_{\rm C}}$ on the disk plane. Animated versions of these plots can be viewed at and downloaded from \color{blue}\smash{http://202.127.29.4/AGN/beiyou/QPO/index\_QPO.html}\color{black}
\label{irradiation_pattern_r90}}
\end{figure}

In order to demonstrate this effect, we plot four snapshots of the irradiation pattern (i.e., incident luminosity $L = \int F_E {\rm d}E$) on the disk in Fig. \ref{irradiation_pattern_r90} for a large truncation radius $R_{\rm tr}=90$, taken at four different values of precession angles, $\gamma/2 \pi=0, 1/4, 1/2, 3/4$.
At the beginning of the precession, $\gamma/2 \pi=0$, i.e., the maximum misalignment between the corona and disk (leading to the maximum illumination of the disk), the irradiation pattern appears to be bright on the left side of the disk ($\sim 180^{\circ}$), but to be faint on the right region ($\sim 0^{\circ}$). The difference in the incident luminosity can be up to about 1.8 orders of magnitude. Meanwhile, the corona is symmetric about the X-axis and the illuminating corona is far away from the BH (i.e., the relativistic effect is negligible); therefore, the irradiation pattern on the left side is roughly symmetric about the X-axis. 
%However, before photons hit the right region of the disk, they will eventuall pass by BH, surrfering from the relativistic effect. That is why the irradiation pattern on the right region of the disk moves anti-clockwise, instead of being symetric about X-axis. 
As the corona precesses counterclockwise, the entire irradiation pattern will globally rotate counterclockwise, as shown in panels (b), (c), and (d). It should be noted that at the middle phase of the precession, i.e., $\gamma/2 \pi=0.5$, when the corona axis is aligned with the disk axis, the irradiation pattern turns out to be symmetric on the disk (panel c), which is expected from the geometrical point of view. 

\begin{figure}
\includegraphics[width=\columnwidth]{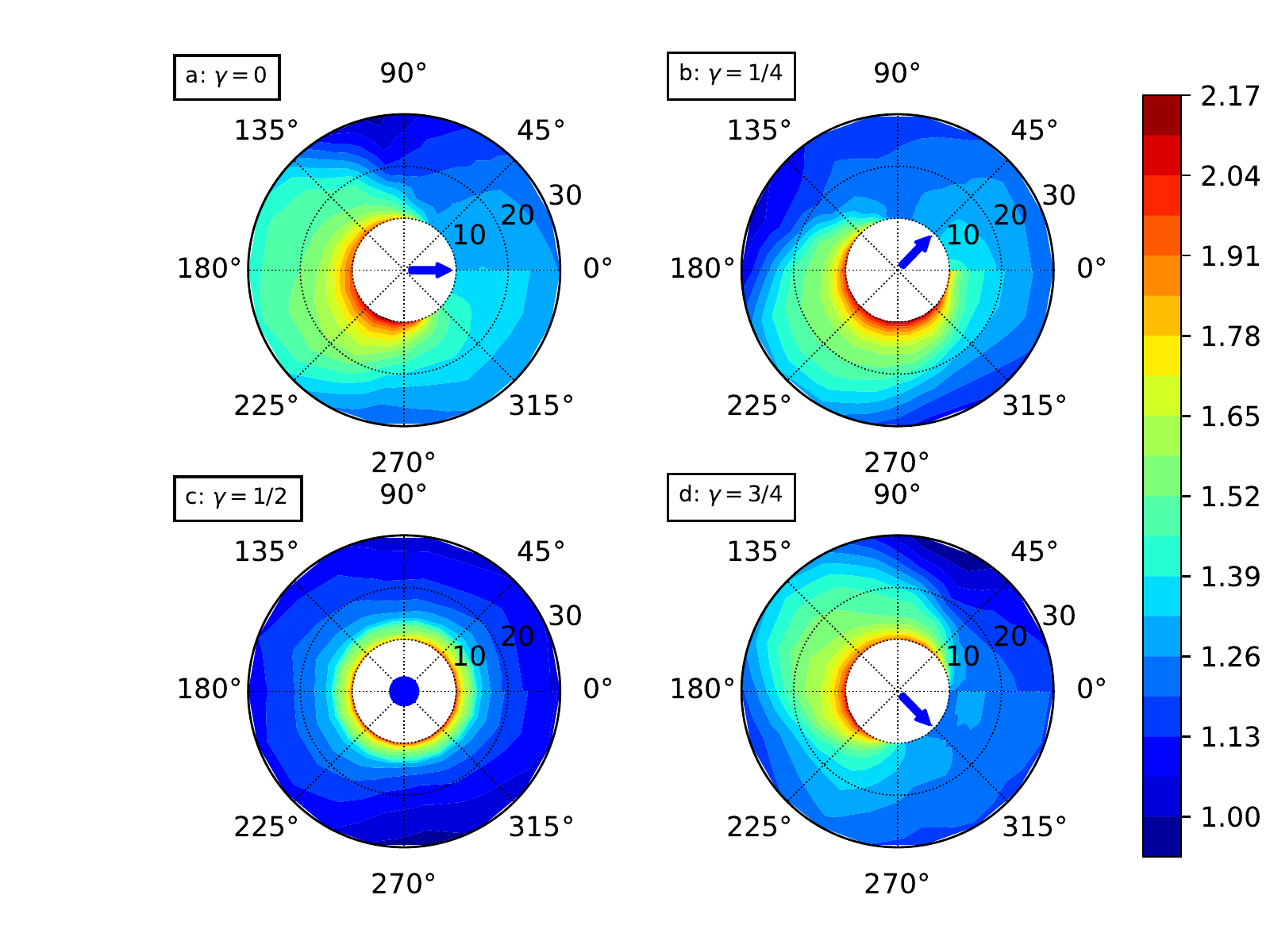}
\caption{
Incident pattern in the disk rest frame, in terms of luminosity $L = \int F_E {\rm d}E$, for the case of the truncated disk plus precessing corona (i.e., Fig. \ref{schematic}). The truncation radius $R_{\rm tr} = 10$. The BH spin $a=0.3$ is assumed. The color bar represents the scaled photon counts (in logarithm) intercepted by the disk. The difference in intercepted luminosity could be up to one order of magnitude. The four panels (a), (b), (c), and (d) correspond to the four precession angles $\gamma/2\pi = 0$, $1/4$, $1/2$, and $3/4$, respectively. In each panel, the blue arrow at the center represents the projection of the corona axis $\bm{J_{\rm C}}$ on the disk plane. Animated versions of these plots can be viewed at and downloaded from \color{blue}\smash{http://202.127.29.4/AGN/beiyou/QPO/index\_QPO.html}\color{black}
\label{irradiation_pattern_r10}}
\end{figure}

In Fig. \ref{irradiation_pattern_r10}, we plot again the irradiation pattern of the precessing corona, but for the small truncation radius $R_{\rm tr}=10$ where photons escaping from the corona are close to the BH. The relativistic effect in this case is significant, compared to the case of the larger truncation radius, so that the irradiation patterns are distorted at all precession phases and are shifted in the azimuthal direction due to the light-bending effect on the incident photons. As the corona precesses, the irradiation pattern on the disk rotates counterclockwise as well.

\begin{figure*}
\includegraphics[width=2\columnwidth]{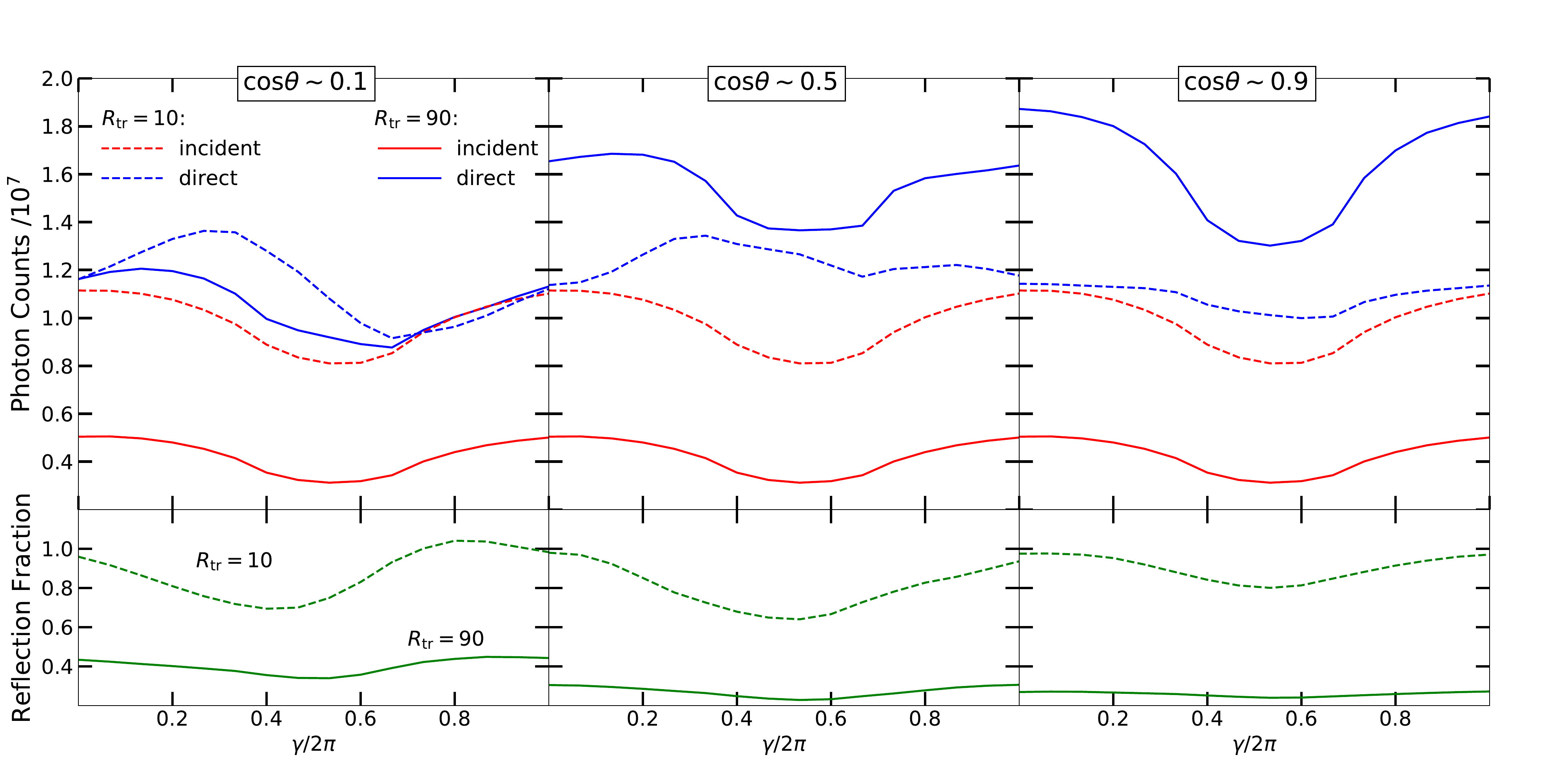} 
\caption{
Upper panels: the normalized photon counts of illuminating the disk (incident; in red color) and being observed by the observer at infinity (direct; in blue color), Bottom panels: the reflection fraction, i.e., the counts ratio of incident photons and direct photons. The solid and dashed lines are for $R_{\rm tr}=90$, and $R_{\rm tr}=10$, respectively. From left to right panels, the inclination angle of the observer corresponds to $\cos \theta \sim 0.1$, 0.5, and 0.9, respectively. The BH spin $a=0.3$ is assumed.
\label{refl_frac}}
\end{figure*}

Because our simulation code can record the information of both the incident flux on the disk and the observed direct flux at infinity, we can study the reflection fraction, i.e., the ratio of the number of photons $\bm{incident}$ on the disk to those going to the observer, which is widely used to infer the relative geometry of the X-ray source with respect to the disk in many codes, e.g. in pexrav (Magdziarz, \& Zdziarski 1995) and relxill (Garc{\'\i}a et al. 2013), also see the discussion in Ingram et al. (2019). In Fig. \ref{refl_frac}, the incident flux (red line) and the observed Comptonization flux (or $\bm{direct}$; blue line) as a function of precession angle/phase are plotted in the upper panels. The resultant reflection fraction (green lines) as a function of precession angle/phase is plotted in the bottom panels. The solid lines are for the large truncation radius $R_{\rm tr}=90$, while the dashed lines are for the small truncation radius $R_{\rm tr}=10$. From the left to right panels, the viewing angles of the observer decreases from $\cos i\sim 0.1$ to  $\cos i\sim 0.9$.
It can be seen that the incident flux shows variation due to the precession, as expected. The minimum of the incident flux occurs at the the middle phase when the corona is aligned with the disk, while the maximum occurs at the the beginning phase when the corona highly tilts up with the maximum misalignment. The variation of the incident flux with the precession phase is independent of the truncation radius. However, as for the observed direct Comptonization flux, that is not the case. It depends on not only the truncation radius, but also the viewing angle. For the truncation radius $R_{\rm tr}=10$, the direct Comptonization flux viewed at a small viewing angle ($\cos i\sim 0.9$; face-on) reaches the minimum at the middle phase and the maximum at the beginning/ending phase.
However, if the viewing angle is large with $\cos i\sim 0.1$ (edge on), the direct Comptonization flux shows the variation in the wave-like form, with the  maximum and minimum fluxes roughly occurring at the first half $\gamma/2\pi \sim 0.3$ and the second half $\gamma/2\pi \sim 0.7$, respectively. In our simulation, the Keplerian rotation of the corona is assumed, which will affect the observed Comptonization flux. At the first half of the precession cycle ($\gamma/2\pi < 0.5$), the approaching side of the corona is facing the observer, so that the Comptonization flux is boosted. At the second half of the precession cycle ($\gamma/2\pi > 0.5$), the receding side of the torus is facing the observer, so that the Comptonization flux is reduced.  

Consequently, the resultant reflection fraction shows variation due to the precession and depends on not only the truncation radius (spectral state) but also on the viewing angle. For large truncation radius, i.e., $R_{\rm tr}=90$, the fraction shows subtle variation with the precession phase, even being constant for large viewing angle. The phase-averaged value roughly ranges from 0.2 to $\sim$ 0.4, which is in the observed range of the reflection fraction of BHXRBs in the low hard state (Zdziarski et al. 1999).
For a small truncation radius, i.e., $R_{\rm tr}=10$ (in this case the spectrum being softened), the fraction also shows variation with the precession phase, but the pattern depends on the viewing angle. At low viewing angle, because the direct Comptonization flux is roughly constant, the reflection fraction almost follows the incident flux, i.e., the minimum occurring at the the middle phase when the corona aligns with the disk, while the maximum occurs at the the beginning phase when the corona highly tilts up with the maximum misalignment. However, at high viewing angle, because the direct Comptonization flux shows the waveform variation, the resultant reflection fraction displays the waveform variation as well, but with opposite sign.
This means that there will be a significant phase difference between the direct Comptonization flux and the reflection fraction when the truncation radius is small.
We also note that the phase-averaged reflection fraction for the small truncation radius, ranging from 0.8 to 1.0, is larger than that for the large truncation radius, which agrees with the observed anticorrelation between the hardness and the reflection fraction (Zdziarski et al. 1999).
 
\subsection{Variation of the reflection}

\begin{figure}
\includegraphics[width=\columnwidth]{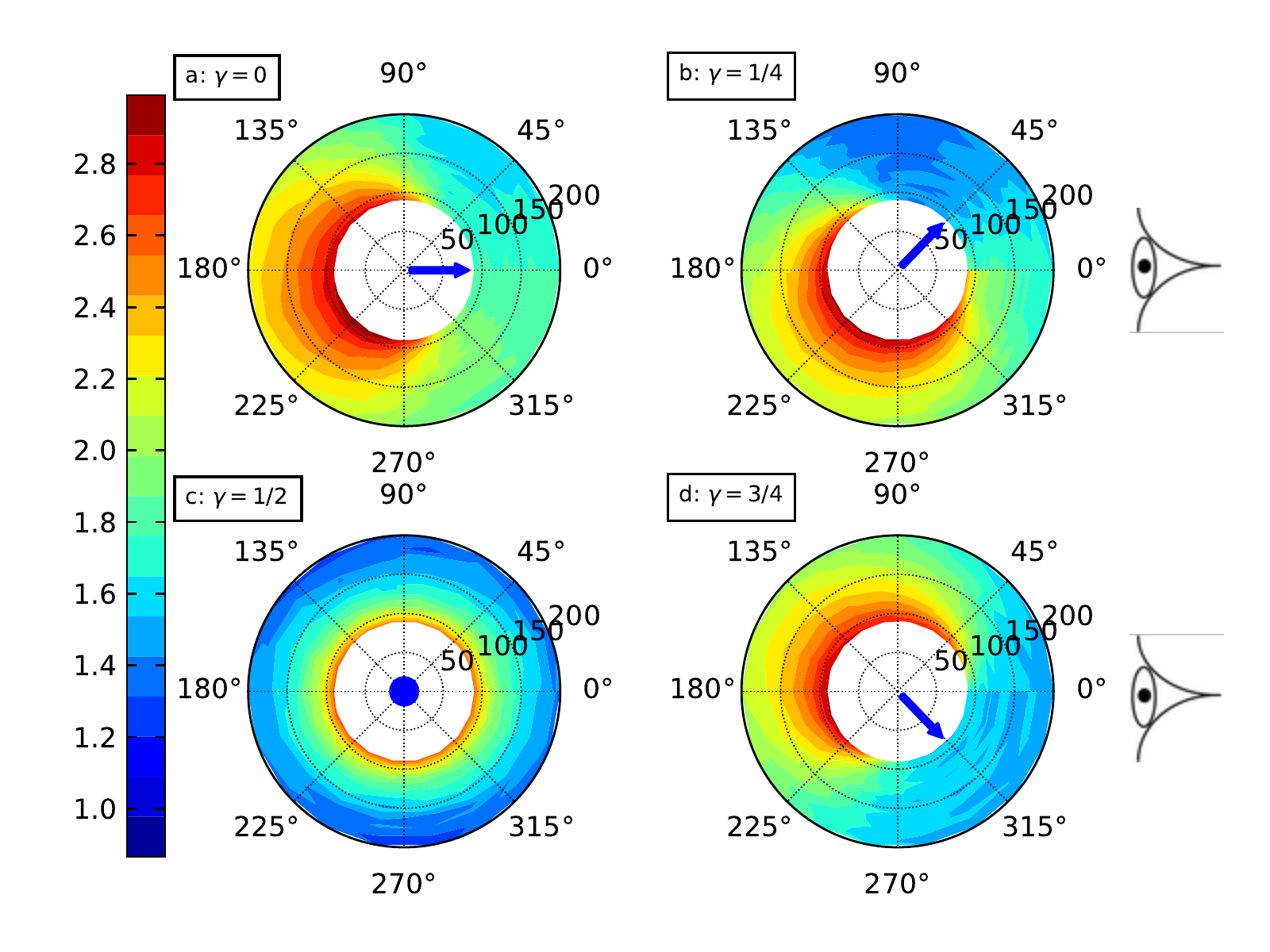} 
\caption{
The reflection pattern in the observer rest frame, in terms of luminosity $L = \int F_E {\rm d}E$, for the truncation radius $R_{\rm tr} = 90$, as seen by an observer at low inclination angle $\cos \theta \sim 0.9$. The BH spin $a=0.3$ is assumed. The color bar represents the scaled luminosity (in logarithm) intercepted by the disk. The four panels (a), (b), (c), and (d) correspond to the four precession angles $\gamma/2\pi = 0$, $1/4$, $1/2$, and $3/4$, respectively. In each panel, the blue arrow at the center represents the projection of the corona axis $\bm{J_{\rm C}}$ on the disk plane. The cartoon of eye indicates the azimuthal position of the observer, $\varphi$ = 0 at infinity.
\label{low_refl_pattern_r90}}
\end{figure}

The corona photons illuminate the disk and will be reprocessed to produce the reflection spectrum including the characteristic Fe K$\alpha$ line.
Fig. \ref{low_refl_pattern_r90} plots the reflection pattern taken at four different values of precession angles, $\gamma/2 \pi=0, 1/4, 1/2, 3/4$, in terms of luminosity $L = \int F_E {\rm d}E$, for the truncation radius $R_{\rm tr} = 90$, as seen by an observer with low inclination angle $\cos \theta \sim 0.9$. It can be seen that the reflection pattern follows the incident patterns.
At the beginning of the precession cycle, $\gamma/2 \pi=0$, i.e., the maximum misalignment between the corona and disk (leading to the maximum illumination of the disk), the reflection pattern appears to be bright on the left side of the disk ($\sim 180^{\circ}$), but to be faint on the right region ($\sim 0^{\circ}$). The difference in the incident luminosity can be up to about 1.8 orders of magnitude. 
As the corona precesses counterclockwise, as a consequence, the entire irradiation pattern will globally counterclockwise rotates, as shown in panel (b), (c), and (d). At the middle phase of the precession, i.e., $\gamma/2\pi=0.5$, when the corona axis is aligned with the disk axis, the reflection pattern turns out to be symmetric on the disk (panel c),

\begin{figure}
\includegraphics[width=\columnwidth]{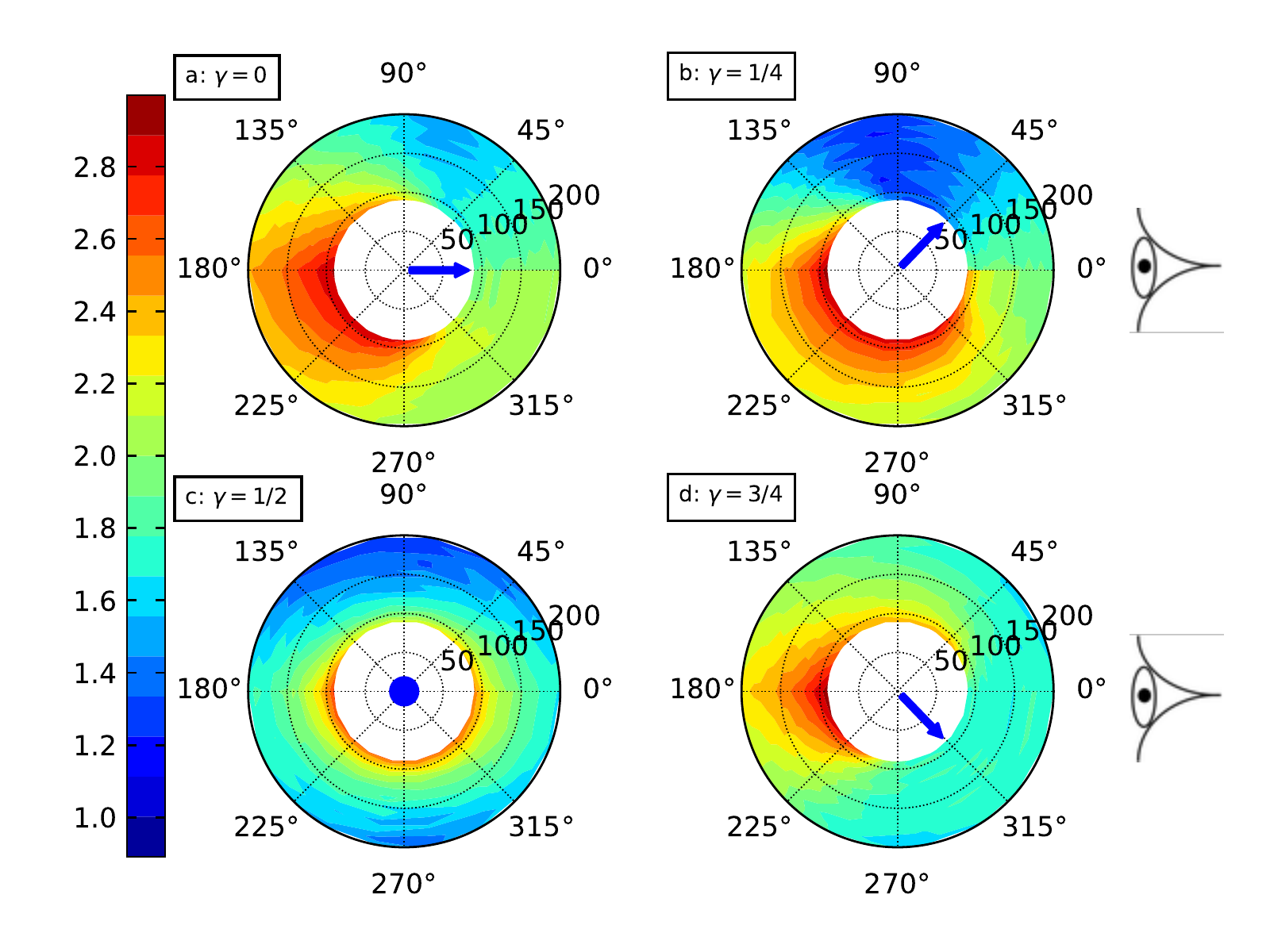} 
\caption{
The reflection pattern in the observer rest frame, in terms of luminosity $L = \int F_E {\rm d}E$, for the truncation radius $R_{\rm tr} = 90$, being observed at high inclination angle $\cos \theta \sim 0.1$. The BH spin $a=0.3$ is assumed. The color bar represents the scaled luminosity (in logarithm) intercepted by the disk. The four panels (a), (b), (c), and (d) correspond to the four precession angles $\gamma/2\pi = 0$, $1/4$, $1/2$, and $3/4$, respectively. In each panel, the blue arrow at the center represents the projection of the corona axis $\bm{J_{\rm C}}$ on the disk plane. The cartoon of eye indicates the azimuthal position of the observer, $\varphi$ = 0 at infinity.
\label{high_refl_pattern_r90}}
\end{figure}

In our simulation, the Keplerian (counterclockwise) rotation of the accretion disk is assumed. When a reprocessed photon leaves the disk, disk rotation will affect not only the observed energy, but also the direction of motion. Therefore, the observed reflection pattern should depend on the inclination angle as well.
In order to demonstrate the effect of the disk rotation on the observed reflection, the reflection pattern at the same truncation radius, but for the high viewing angle, is plotted in Fig. \ref{high_refl_pattern_r90}.
Compared to the face-on case, as for the observed reflection pattern in the edge-on case, the observed reflection from the receding region of the disk is reduced, while the observed reflection from the approaching region of the disk is boosted.

%\begin{figure}
%\includegraphics[width=\columnwidth]{./paper_plot/phi_0_low_reflection_flux_t15_b15_2p15_r10_a0p3.pdf} 
%\caption{
%reflection pattern with small truncation radius $R_{\rm tr} = 10$ and low viewing angle
%\label{low_refl_pattern_r10}}
%\end{figure}

\begin{figure}
\includegraphics[width=\columnwidth]{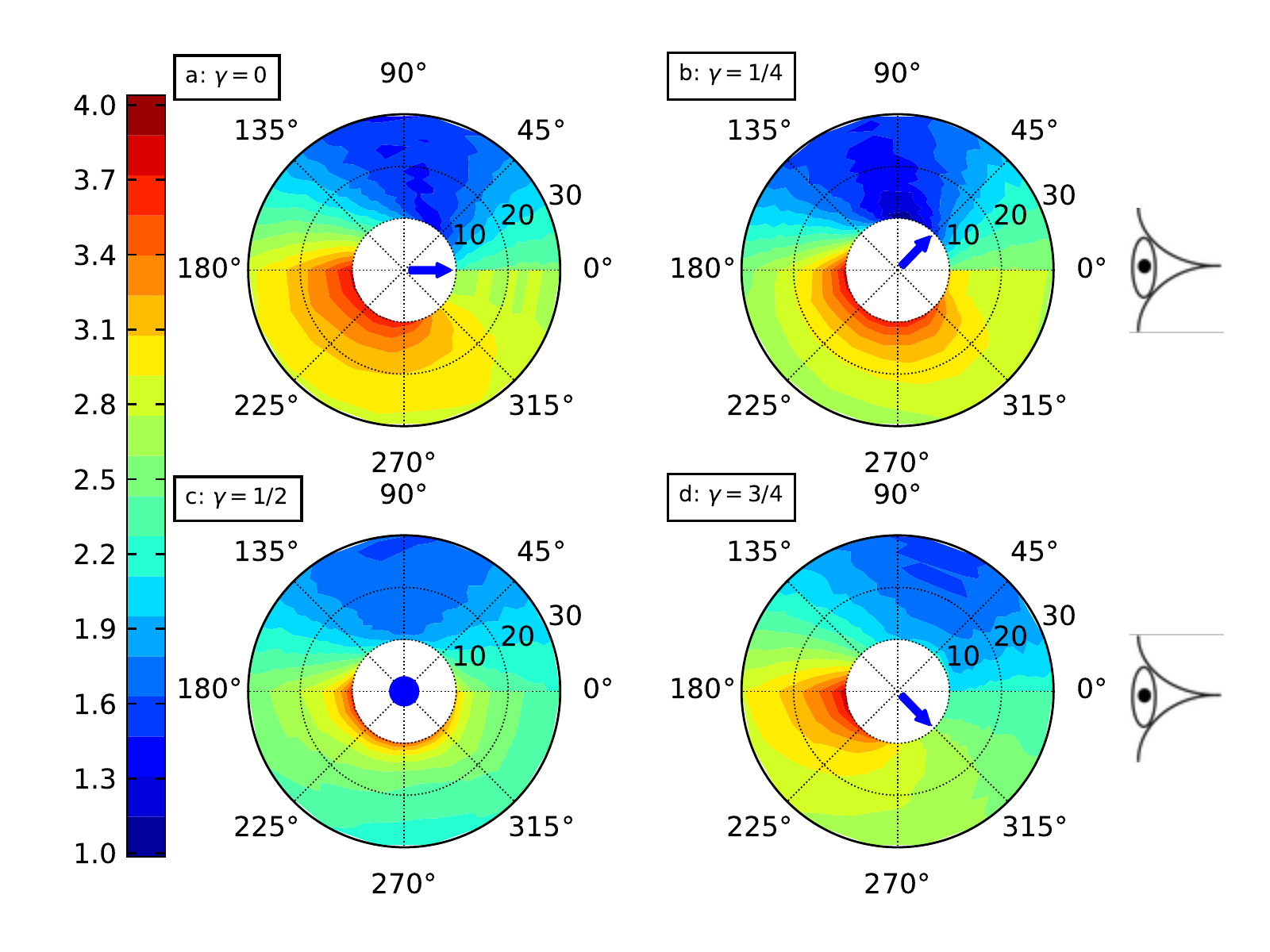} 
\caption{
The reflection pattern in the observer rest frame, in terms of luminosity $L = \int F_E {\rm d}E$, for the truncation radius $R_{\rm tr} = 10$, being observed at high inclination angle $\cos \theta \sim 0.1$. The BH spin $a=0.3$ is assumed. The color bar represents the scaled luminosity (in logarithm) intercepted by the disk. The four panels (a), (b), (c), and (d) correspond to the four precession angles $\gamma/2\pi = 0$, $1/4$, $1/2$, and $3/4$, respectively. In each panel, the blue arrow at the center represents the projection of the corona axis $\bm{J_{\rm C}}$ on the disk plane. The cartoon of eye indicates the azimuthal position of the observer, $\varphi$ = 0 at infinity.
\label{high_refl_pattern_r10}}
\end{figure}

The simulations above are for the large truncation radius, which means the reflecting photons are far away from the BH. If the disk is truncated at small radius, close to the BH, e.g., $R_{\rm tr} = 10$, then the reflecting photons will inevitably suffer from relativistic effects (e.g., light bending), which will significantly affect the photon trajectory to infinity. In Fig. \ref{high_refl_pattern_r10}, we plot the reflection pattern for the small truncation radius $R_{\rm tr} = 10$ and high viewing angle $\cos \theta \sim 0.1$. Differing from the case of large truncation radius $R_{\rm tr} = 90$, as the corona precesses, the brightest reflection keeps coming from the approaching side of the disk, while the dimmest reflection keeps coming from the receding side of the disk. In other words, the receding side of the disk never dominates over the approaching side of the disk in the observed reflection flux.

%(我们程序未来可以做的)Note that large value of h/r gives a reasonable reflection fraction
%but underpredicts the QPO rms. If we had considered, for example,
%an overlap region between disc and flow, disc flares or a small
%disc scale height, we could have achieved a reasonable reflection
%fraction and the correct QPO rms (for this we would also need to
%consider the variation in disc seed photons) for a far lower value of
%h/r. However, these effects are all very difficult to model and our
%assumed geometry should not significantly affect the final results.
%Thus we choose the fiducial parameter values to give reasonable
%results for a simplified geometry.

\subsection{Variation of the Fe K$\alpha$ line}
The reflection patterns shown above could also represent the contribution to the overall luminosity of the Fe K$\alpha$ line from different regions of the disk. Because the shift in photon energy due to both gravitational and Doppler effects depends on the emitting location on the disk, the overall spectroscopic profile of the Fe K$\alpha$ line is determined by the reflection pattern.  
Because the reflection pattern varies as the corona precesses and the variation depends on not only the truncation radius but also the inclination angle, the overall spectroscopic shape of the Fe K$\alpha$ line should depend on the truncation radius and the inclination angle as well.

\begin{figure}
\includegraphics[width=\columnwidth]{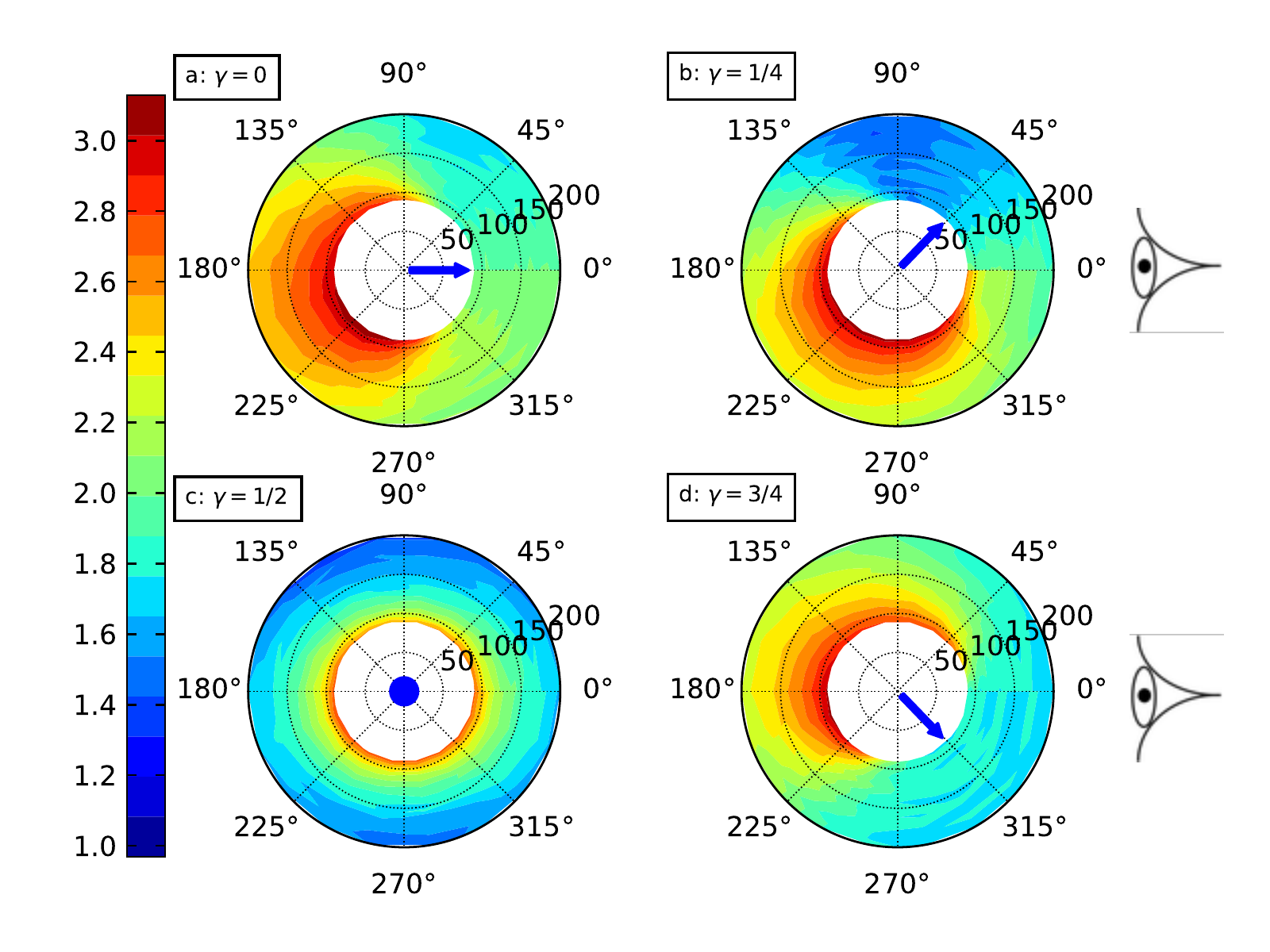} 
\caption{
The reflection pattern in the observer rest frame, in terms of luminosity $L = \int F_E {\rm d}E$, for the truncation radius $R_{\rm tr} = 90$, being observed at middle inclination angle $\cos \theta \sim 0.5$. The BH spin $a=0.3$ is assumed. The color bar represents the scaled luminosity (in logarithm) intercepted by the disk. The four panels (a), (b), (c), and (d) correspond to the four precession angles $\gamma/2\pi = 0$, $1/4$, $1/2$, and $3/4$, respectively. In each panel, the blue arrow at the center represents the projection of the corona axis $\bm{J_{\rm C}}$ on the disk plane. The cartoon of eye indicates the azimuthal position of the observer $\varphi$ = 0 at infinity. Animated versions of these plots can be viewed at and downloaded from \color{blue}\smash{http://202.127.29.4/AGN/beiyou/QPO/index\_QPO.html}\color{black}
\label{mid_refl_pattern_r90}}
\end{figure}

\begin{figure}
\includegraphics[width=\columnwidth]{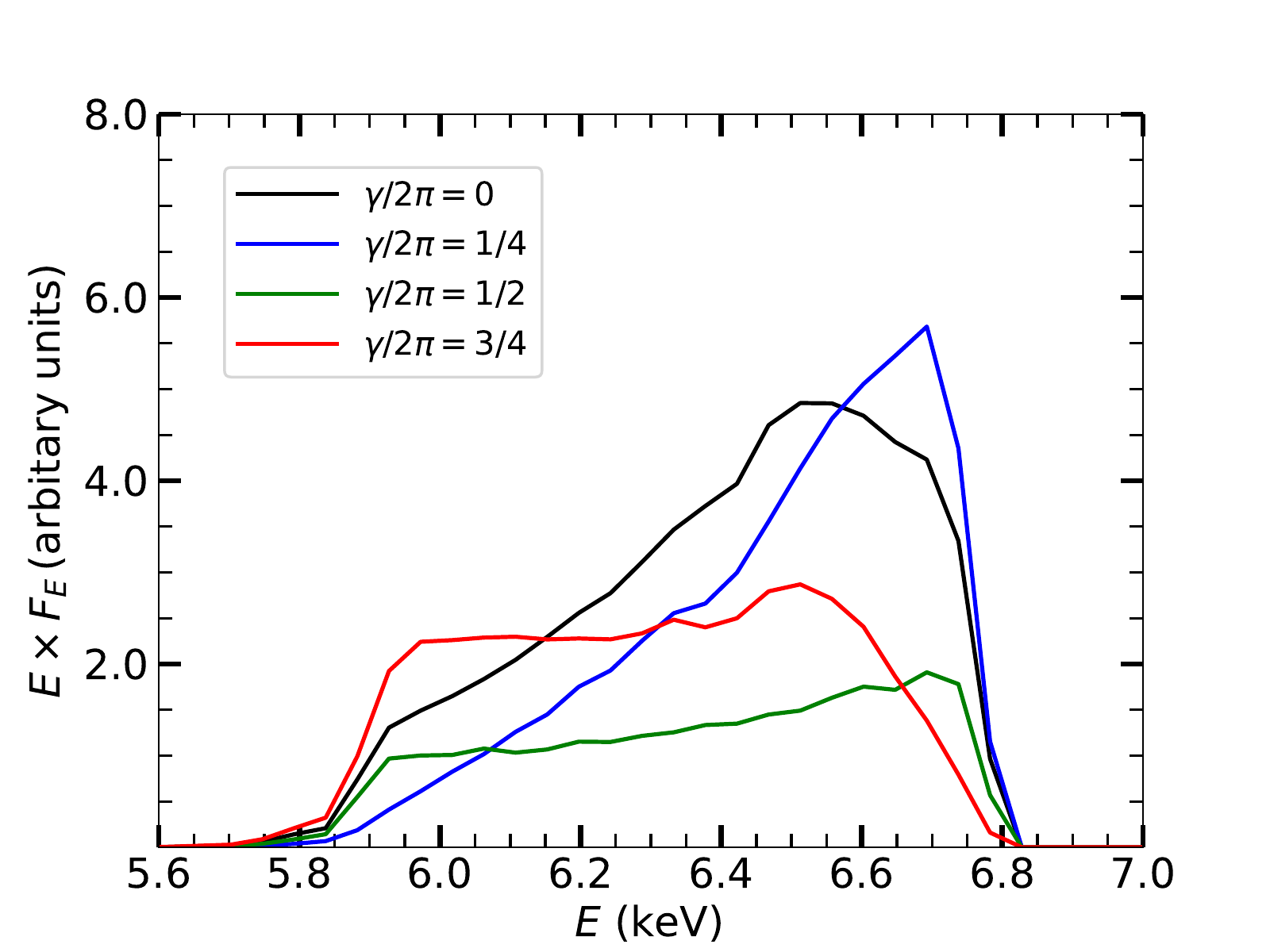}
\caption{
The observed Fe K$\alpha$ line for large truncation radius $R_{\rm tr} = 90$ and middle inclination angle $\cos \theta \sim 0.5$. The black, blue, green and red profiles correspond to four specific precession angles, $\gamma/2\pi = 0$, $1/4$, $1/2$, and $3/4$, respectively. The BH spin $a=0.3$ is assumed. Animated versions of these plots can be viewed at and downloaded from \color{blue}\smash{http://202.127.29.4/AGN/beiyou/QPO/index\_QPO.html}\color{black}
\label{line_shape_r90}}
\end{figure}

In Fig. \ref{mid_refl_pattern_r90}, we plot the reflection pattern for large truncation radius $R_{\rm tr}=90$ as seen by an observer at the middle viewing angle $\cos \theta \sim 0.5$. The corresponding spectroscopic profile of the Fe K$\alpha$ line is plotted in Fig. \ref{line_shape_r90}. At the beginning of the precession cycle, the brightest reflection primarily occurs at the region hidden from the observer (at the azimuth of $\sim 180^\circ$ on the disk), where the Doppler effect is not strong (from the point of view of the observer). Besides, the brightest reflection also partially occupies the approaching side of the disk. Consequently, the overall peak of the Fe K$\alpha$ line is subtly shifted to $\sim 6.5\, \rm keV$ (the black line). As the reflection pattern rotates counterclockwise to the phase $\gamma/2\pi = 0.25$, the reflection from the approaching side of the disk is enhanced, thus the Fe K$\alpha$ line will be dominated by the approaching disk and its peak energy is shifted to higher energy $\sim 6.7\, \rm keV$ (the blue line). At the middle phase of the precession, i.e., $\gamma/2\pi = 0.5$, when the corona is aligned with the disk, the overall reflection strength (luminosity) is dimmest. The reflection pattern from the approaching side is slightly enhanced due to the disk rotation, which leads to the blue wing of the Fe K$\alpha$ line being slightly higher than the red wing (the green line). As precession continues, e.g., $\gamma/2\pi = 0.75$, when the reflection pattern from the receding side of the disk dominates over that from the approaching side of the disk, we see that the high-energy tail of the Fe K$\alpha$ line is reduced (red line), in comparison to the case of $\gamma/2\pi = 0.5$ (green line).

\begin{figure}
\includegraphics[width=\columnwidth]{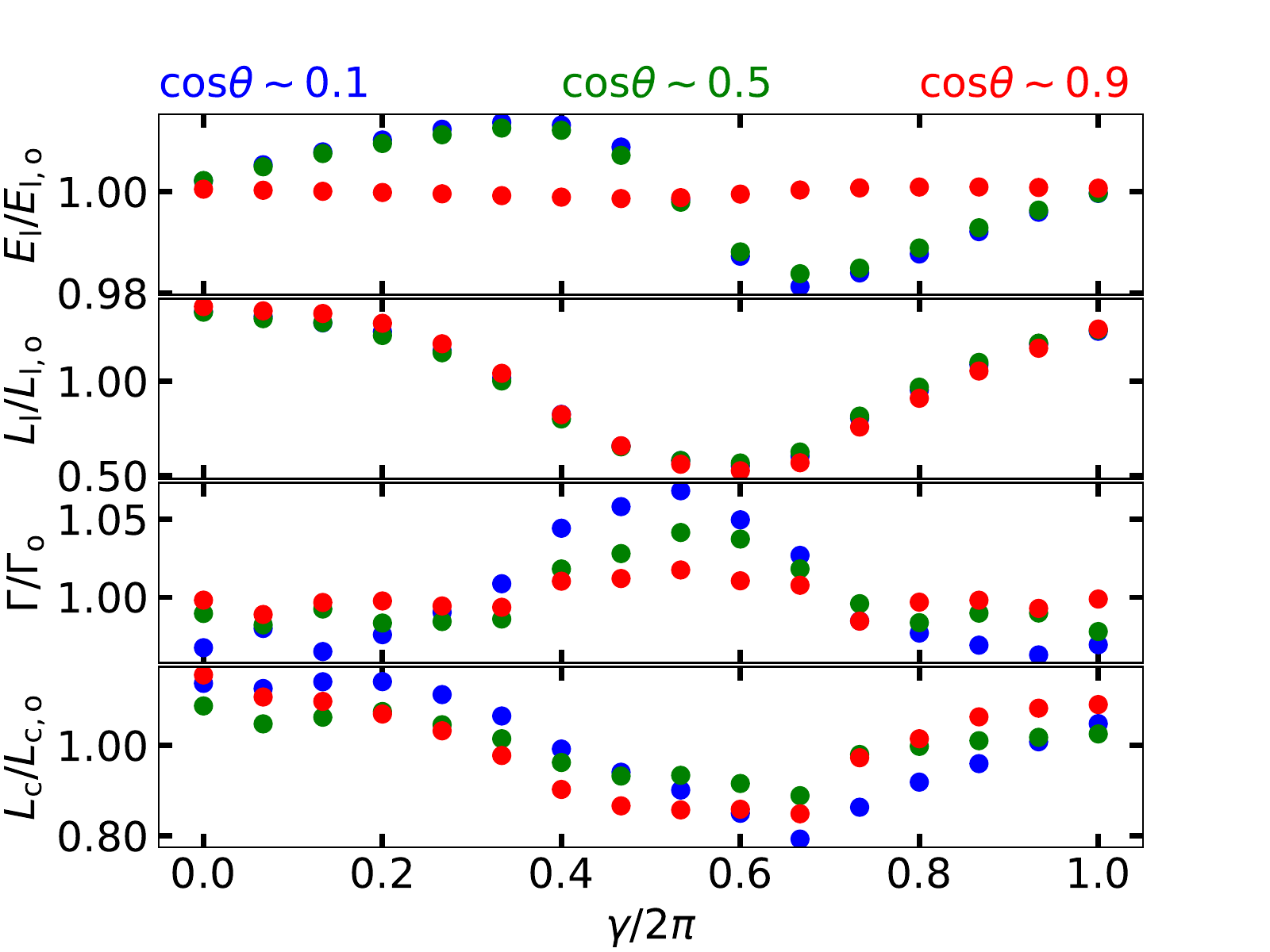}
\caption{
The spectral properties of the observed 2-10 keV radiation, as a function of precession angle $\gamma/2\pi$, for large truncation radius $R_{\rm tr} = 90$. From top to bottom, each panel corresponds to the central energy of the line $E_{\rm l}$, the iron line luminosity $L_{\rm l}$, the spectral slope $\Gamma$, and continuum flux $L_{\rm C}$. Correspondingly, the phase-averaged values are labeled as $E_{\rm l,o}$, $L_{\rm l,o}$, $\Gamma_{\rm o}$ and $L_{\rm C,o}$. 
%In order to demonstrate the variation of these spectral parameters, the y-axis in all panels represent the ratio of the phase-dependent value and the phase-averaged value, i.e., $E_{\rm l}/E_{\rm l,o}$, $L_{\rm l}/L_{\rm l,o}$, $\Gamma/\Gamma_{\rm o}$ and $L_{\rm c}/L_{\rm c,o}$.
The blue, green, and red points are for the viewing angles $\cos \theta \sim 0.1$, 0.5, and 0.9, respectively. The BH spin $a=0.3$ is assumed.
\label{line_para_r90}}
\end{figure}

In order to quantitatively demonstrate the variation of the Fe K$\alpha$ line, we define the flux-weighted central energy of the line $E_{\rm l}$, as follows:
\begin{equation}\label{central_energy}
   E_{\rm l} = \frac{\int\,EF(E)\,{\rm d}E}{\int\,F(E)\,{\rm d}E}  
\end{equation}

We plot the flux-weighted centroid energy of the Fe K$\alpha$ line $E_{\rm l}$ as a function of precession angle $\gamma/2\pi$, for large truncation radius $R_{\rm tr} = 90$, in Fig. \ref{line_para_r90}. Also plotted are the line luminosity $L_{\rm l}$, the spectral slope $\Gamma$, and the continuum flux $L_{\rm C}$ between 2 and 10 keV. For clarity, these values are plotted as the ratio between the phase-dependent values and the phase-averaged values. It can be seen that the line flux $L_{\rm l}$ and the continuum flux $L_{\rm C}$ between 2 and 10 keV, with respect to the phase-averaged values, show large variation by up to 50\% and 20\%, respectively. The variation of these two spectral parameters with the precession phase is insensitive to the inclination angle. The spectral slope $\Gamma$ also varies with precession phase by about $\sim 5\%$.
The spectrum becomes softest at the precession angle $\gamma/2\pi \sim0.5$ when the corona is aligned with the disk axis. We note that the centroid energy $E_{\rm l}$ varies with precession phase by up to $\sim $2\%, for large inclination angles (blue and green lines), being dominated by the blueshift effect at the first half of the precession period while being dominated by the redshift effect at the second half of the precession period. The centroid energy was observed to systematically vary in H1743-322 (ID17), although the observed variation pattern is complex in comparison to the prediction here. For low inclination angle (red line), the predicted variation pattern is almost constant. Given the dependence of these spectral properties on the inclination in Fig. \ref{line_para_r90}, the variation of the centroid energy $E_{c}$ with the phase could be a good diagnostic for whether the inclination angle is low or large.

\begin{figure}
\includegraphics[width=\columnwidth]{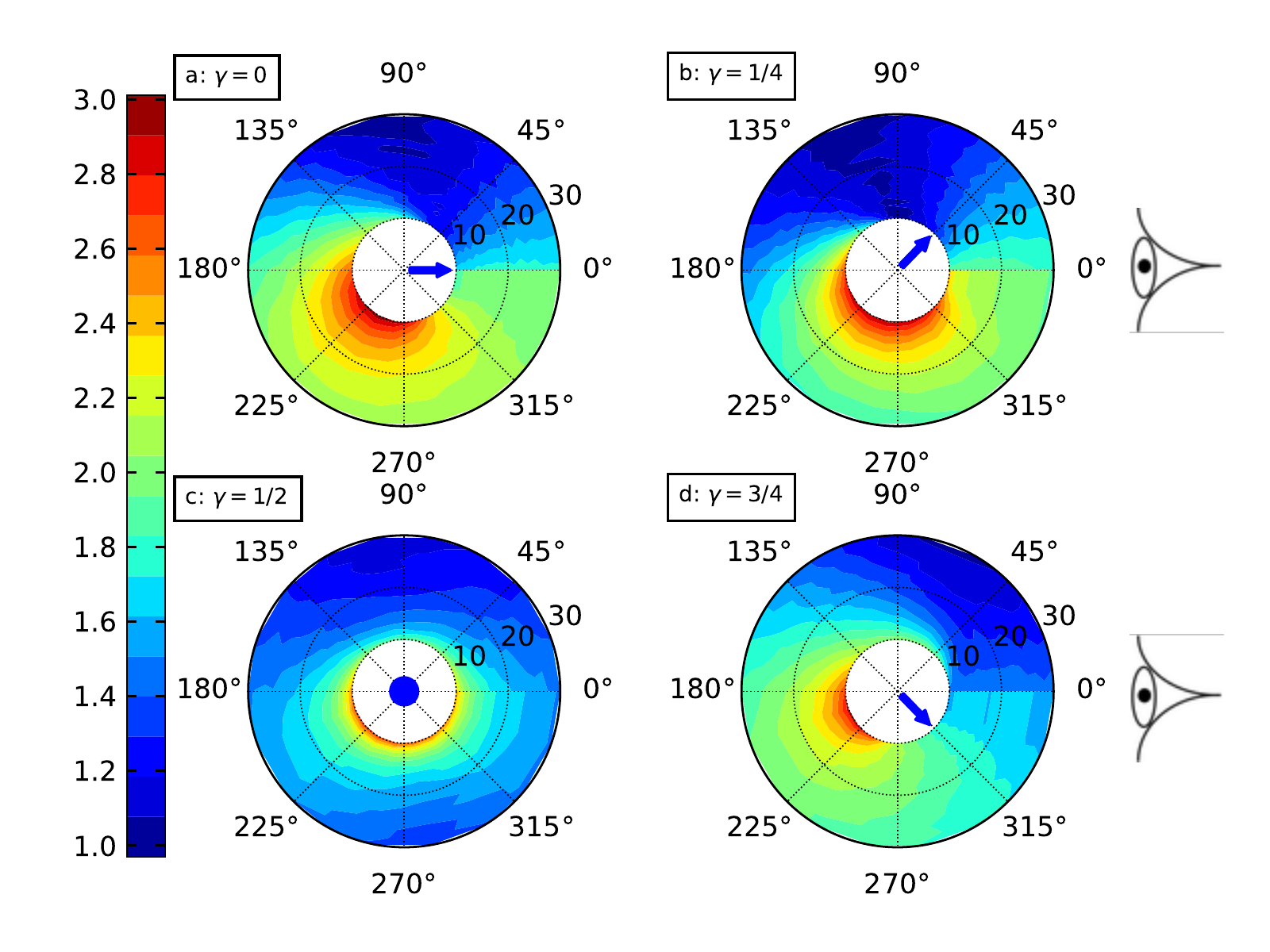}
\caption{
The reflection pattern in the observer rest frame, in terms of luminosity $L = \int F_E {\rm d}E$, for the truncation radius $R_{\rm tr} = 10$, being observed at middle inclination angle $\cos \theta \sim 0.5$. The BH spin $a=0.3$ is assumed. The color bar represents the scaled luminosity (in logarithm) intercepted by the disk. The four panels (a), (b), (c), and (d) correspond to the four precession angles $\gamma/2\pi = 0$, $1/4$, $1/2$, and $3/4$, respectively. In each panel, the blue arrow at the center represents the projection of the corona axis $\bm{J_{\rm C}}$ on the disk plane. The cartoon of eye indicates the azimuthal position of observer $\varphi = 0$ at infinity. Animated versions of these plots can be viewed at and downloaded from \color{blue}\smash{http://202.127.29.4/AGN/beiyou/QPO/index\_QPO.html}\color{black}
\label{mid_refl_pattern_r10}}
\end{figure}

\begin{figure}
\includegraphics[width=\columnwidth]{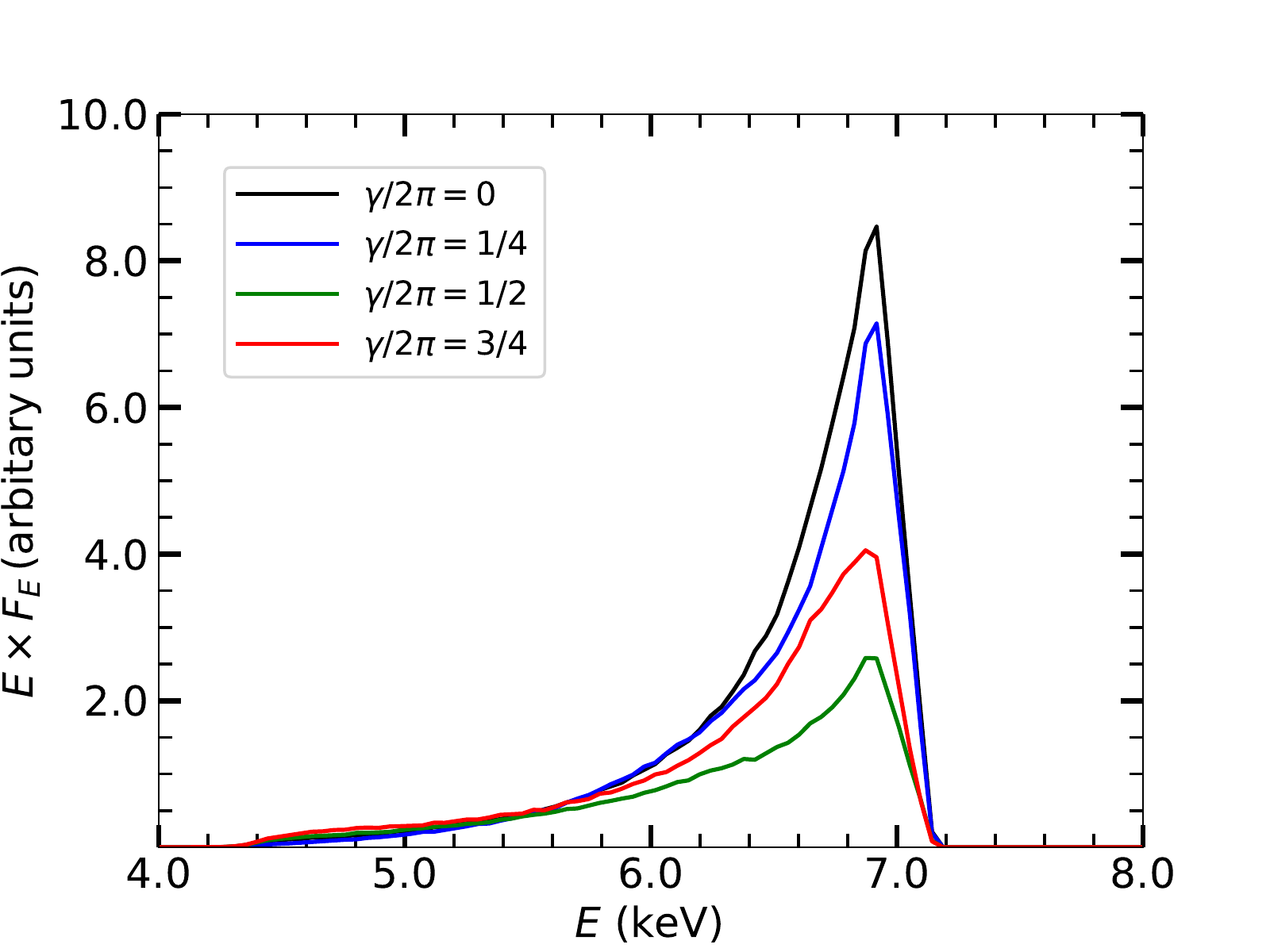} 
\caption{
The observed Fe K$\alpha$ line for large truncation radius $R_{\rm tr} = 10$ and middle inclination angle $\cos \theta \sim 0.5$. The black, blue, green, and red profiles correspond to four specific precession angles, $\gamma/2\pi = 0$, $1/4$, $1/2$, and $3/4$, respectively. The BH spin $a=0.3$ is assumed. Animated versions of these plots can be viewed at and downloaded from \color{blue}\smash{http://202.127.29.4/AGN/beiyou/QPO/index\_QPO.html}\color{black}
\label{line_shape_r10}}
\end{figure}

In Fig. \ref{mid_refl_pattern_r10}, we plot the reflection patterns taken at four different precession angles $\gamma/2\pi = 0$, $1/4$, $1/2$, and $3/4$ for small truncation radius $R_{\rm tr}=10$, as seen by an observer at the middle viewing angle $\cos \theta \sim 0.5$. The corresponding spectroscopic shape of the Fe K$\alpha$ line is plotted in Fig. \ref{line_shape_r10}. Differing from the case of the larger truncation radius, in the case of the smaller truncation radius, the brightest reflection pattern constantly comes from the approaching side of the disk, while the dimmest reflection constantly comes from the receding side of the disk. The entire reflection is dominated by the approaching side of the disk. Therefore, over the full precession cycle, the corresponding Fe K$\alpha$ line always peaks at the higher energy $E\sim$ 6.8 keV due to Doppler shift and is skewed to low energy $E\sim$ 4 keV. The main difference between the four Fe K$\alpha$ line profiles in Fig. \ref{line_shape_r10} is the peak luminosity.

\begin{figure}
\includegraphics[width=\columnwidth]{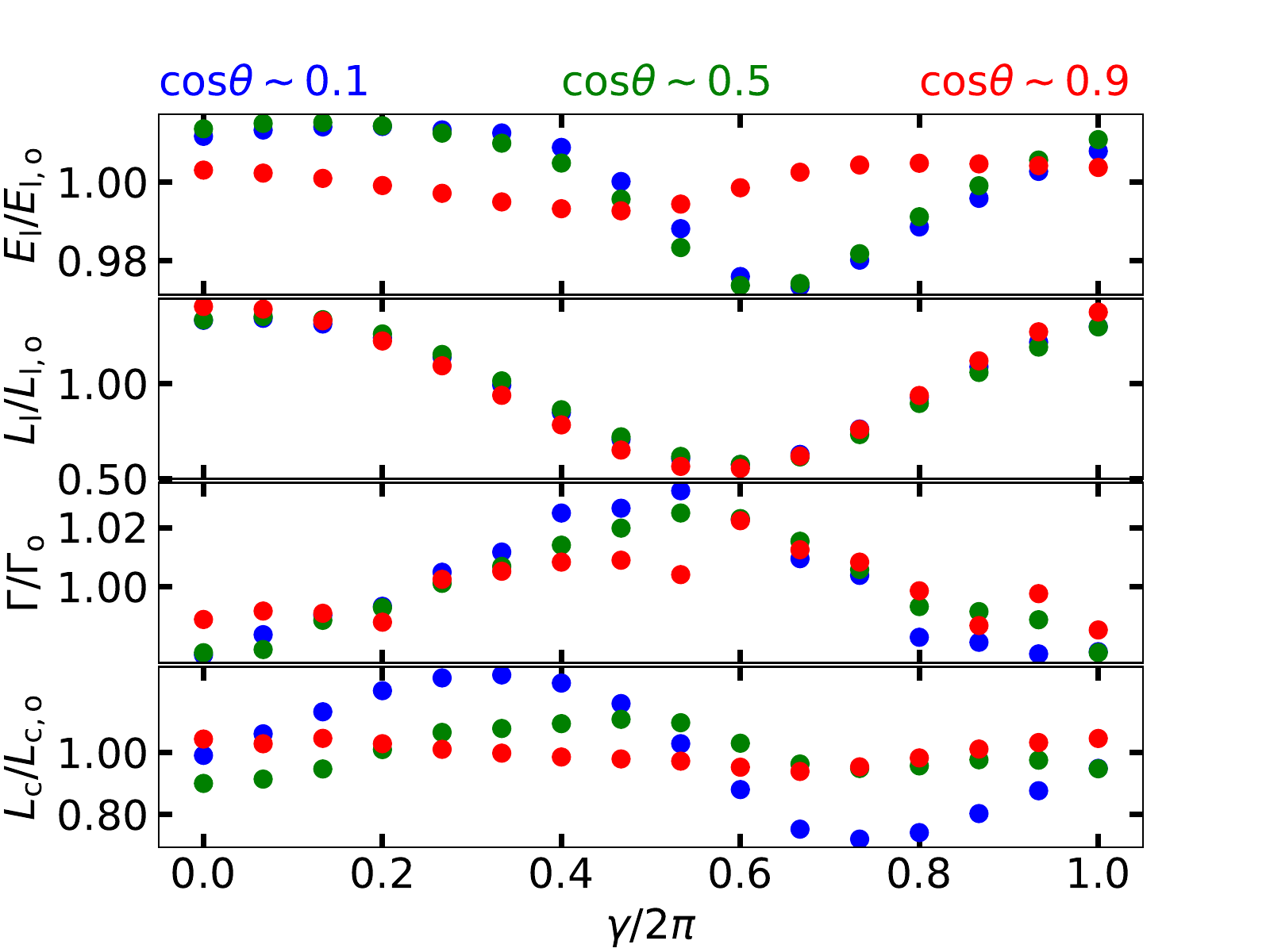} 
\caption{
The spectral properties of the observed 2-10 keV radiation, as a function of precession angle $\gamma/2\pi$, for large truncation radius $R_{\rm tr} = 10$. From top to bottom, each panel corresponds to the central energy of the line $E_{\rm l}$, the iron line flux $L_{\rm l}$, the spectral slope $\Gamma$, and continuum flux $L_{\rm C}$. Correspondingly, the phase-averaged values are labeled as $E_{\rm l,o}$, $L_{\rm l,o}$, $\Gamma_{\rm o}$ and $L_{\rm C,o}$. 
%In order to demonstrate the variation of these spectral parameters, the y-axis in all panels represent the ratio of the phase-dependent value and the phase-averaged value, i.e., $E_{\rm l}/E_{\rm l,o}$, $L_{\rm l}/L_{\rm l,o}$, $\Gamma/\Gamma_{\rm o}$ and $L_{\rm c}/L_{\rm c,o}$.
The blue, green, and red points are for the viewing angles $\cos \theta \sim 0.1$, 0.5, and 0.9, respectively. The BH spin $a=0.3$ is assumed.
\label{line_para_r10}}
\end{figure}

%The spectral properties between 2-10 keV, i.e., the emission line central energy $E_{c}$, line flux $F_{\rm line}$, the spectral slope $\Gamma$, and the continuum flux $F_{2-10}$, for different precession phase is plotted in Fig. \ref{line_para_r10_phi_2}. For clarity, these values are plotted as the ratio of the phased-dependent values and the phase-averaged values. It can be seen that, the central energy $E_{c}$ and the continuum flux $F_{2-10}$ shows variation, reaching up to $\sim 2$\%, and $\sim$ 20 \%, for large inclination angles (blue and green lines), but is less variable for low inclination angle (red line). The line flux $F_{\rm line}$, the spectral slope $\Gamma$ can vary up to $\sim 30$\%, $\sim 3$\%, but are insensitive to the inclination angle. From the perspective of the inclination dependence of these spectral properties, the variation of central energy $E_{c}$ and the continuum flux $F_{2-10}$ could be good diagnostic to the incination angle.  

We plot the flux-weighted centroid energy of the Fe K$\alpha$ line $E_{\rm l}$, line luminosity $L_{\rm l}$, the spectral slope $\Gamma$, and the continuum flux $L_{\rm C}$ between 2 and 10 keV, as a function of precession angle $\gamma/2\pi$, for small truncation radius $R_{\rm tr} = 10$, in Fig. \ref{line_para_r10}. Again, these values are plotted as the ratio of the phase-dependent values to the phase-averaged values. Although the truncation radius decreases from  $R_{\rm tr} = 90$ to $R_{\rm tr} = 10$, the variation pattern of the Fe K$\alpha$ line $E_{\rm l}$, line luminosity $L_{\rm l}$ and the spectral slope $\Gamma$ with the precession phase in Fig. \ref{line_para_r90} and \ref{line_para_r10}, are not changed. We note that in the case of $R_{\rm tr} = 10$, the variation pattern of the continuum flux $L_{\rm C}$ with precession phase depends on the inclination angle, i.e., showing modulation by up to $\sim 20$\% for large inclination angles $\cos \theta \sim 0.1$ (blue points), but is less variable for low inclination angle $\cos \theta \sim 0.9$ (red points). From this perspective, the phase-dependent continuum flux $L_{\rm C}$ could be a good diagnostic for the evolution of the truncation geometry, if the source is viewed with low inclination angle.  

%
%We also repeat the simulation above for the large black hole spin $a=0.9$. It was found that there is not significant difference between the small and large black hole spin, in terms of both incident/reflection pattern and reflection fraction, except the enhancement in the luminosity (the dashed lines). 

\section{Discussion}

In You et al. (2018), the effect of the parameters, e.g., the accretion geometry (i.e., the truncation radius of the disk), the observer position (in terms of inclination angle and azimuthal angle), etc., on the observed QPO properties (e.g., energy-dependent fractional variability amplitude) of the continuum emission was studied in the scenario of Lense-Thirring precession. In this work, we mainly studied the phase-resolved irradiation/reflection (including the Fe K$\alpha$ line) and the resultant spectral properties, in the same scenario of precession.
This will allow us to further study the phase lag between these spectral properties.

\subsection{Phase lag}
%The observed modulation of the X-ray flux (QPO) in BHXRBs might be caused by the Lense-Thirring precession if considering the inclination effect, e.g., the wobbling of the projected area of the X-ray source. This effect could be additionally enhanced if the Doppler rotation/boosting and the relativistic effect are taken into account, which requires the X-ray source is close to BH.
In Fig. \ref{line_para_r90} and \ref{line_para_r10}, the luminosity of the Fe K$\alpha$ line $L_{\rm l}$ is modulated by the precession phase. But, the variation pattern of the line luminosity relative to the phase-averaged value is independent of the inclination angle and the truncation radius (i.e., corresponding to the spectral state in the truncation model). This means that the Fe K$\alpha$ line luminosity $L_{\rm l}$ could be taken as a good reference to study the phase lag of other spectral properties with respect to it. As the truncation radius decreases from $R_{\rm tr}=90$ to 10, there is no obvious phase lag for the spectral slope $\Gamma$. For a large inclination angle, the minimum of the centroid energy of the Fe K$\alpha$ line $E_{\rm l}$ is roughly at the same phase $\gamma/2\pi \sim 0.65$ in the cases of $R_{\rm tr}=90$ and 10, although the line profiles and their variation with the precession phase are absolutely different for different truncation radii. At the truncation radius $R_{\rm tr}=90$, we found that there is no significant phase lag between the continuum flux $L_{\rm C}$ and the Fe K$\alpha$ line luminosity $L_{\rm l}$, when the inclination angle is small with $\cos \theta \sim 0.9$. However, when the inclination angle is large, e.g., $\cos \theta \sim 0.1$, the continuum flux $L_{\rm C}$ slightly lags the Fe K$\alpha$ line luminosity $L_{\rm l}$. This phase lag is caused by the Doppler effect due to the rotation of the corona in the case of high inclination angle. This effect could be more significant for the faster rotation of the corona when the truncation radius (i.e., the outer radius of the corona) is small, $R_{\rm tr}=10$.
Therefore, the continuum flux $L_{\rm C}$ apparently lags the luminosity of the Fe K$\alpha$ line $L_{\rm l}$.

\subsection{Effect of the observer azimuth}
The simulation results above are for the case of the observer with the azimuthal position of $\varphi=0$, i.e., the precessing corona is directed toward the observer. However, the azimuth $\varphi$ of the observer is unknown. It has been shown that the QPO properties (variability amplitude and polarization) can be very different for different azimuthal postions of the observer (Ingram et al. 2015; You et al. 2018). In order to study the effect of the observer azimuth on the phase-resolved emission, in this section we consider the case of the observer being behind the precessing torus, i.e., the azimuthal position of $\varphi=\pi$. 

\begin{figure}
\includegraphics[width=\columnwidth]{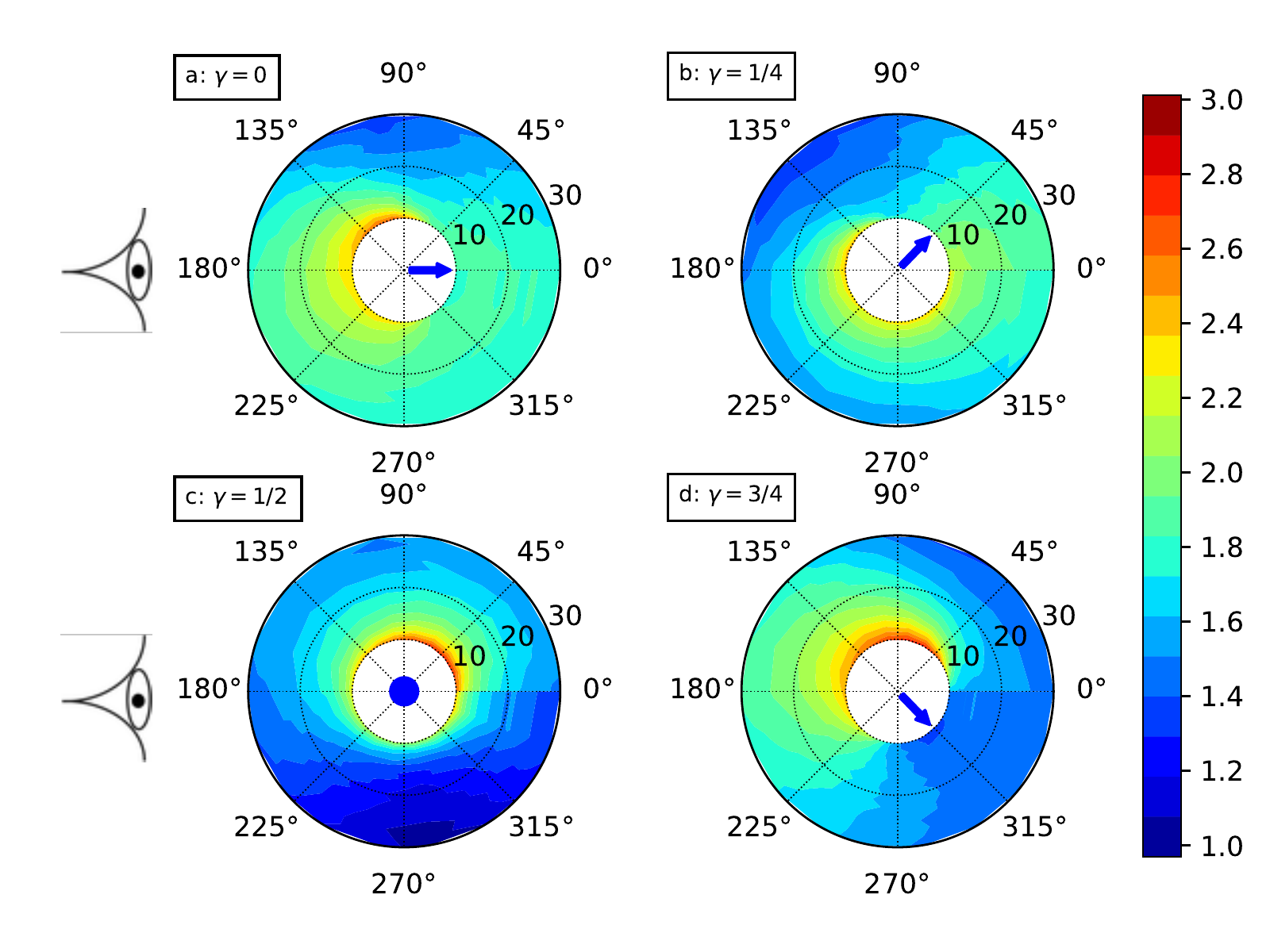} 
\caption{
The reflection pattern in the observer rest frame, in terms of luminosity $L = \int F_E {\rm d}E$, for the truncation radius $R_{\rm tr} = 10$, being observed at middle inclination angle $\cos \theta \sim 0.5$. The color bar represents the scaled luminosity (in logarithm) intercepted by the disk. The four panels (a), (b), (c), and (d) correspond to the four precession angles $\gamma/2\pi = 0$, $1/4$, $1/2$, and $3/4$, respectively. The azimuthal angle of the observer $\varphi = \pi$. The BH spin $a=0.3$ is assumed.
\label{mid_refl_pattern_r10_phi2}}
\end{figure}

\begin{figure}
\includegraphics[width=\columnwidth]{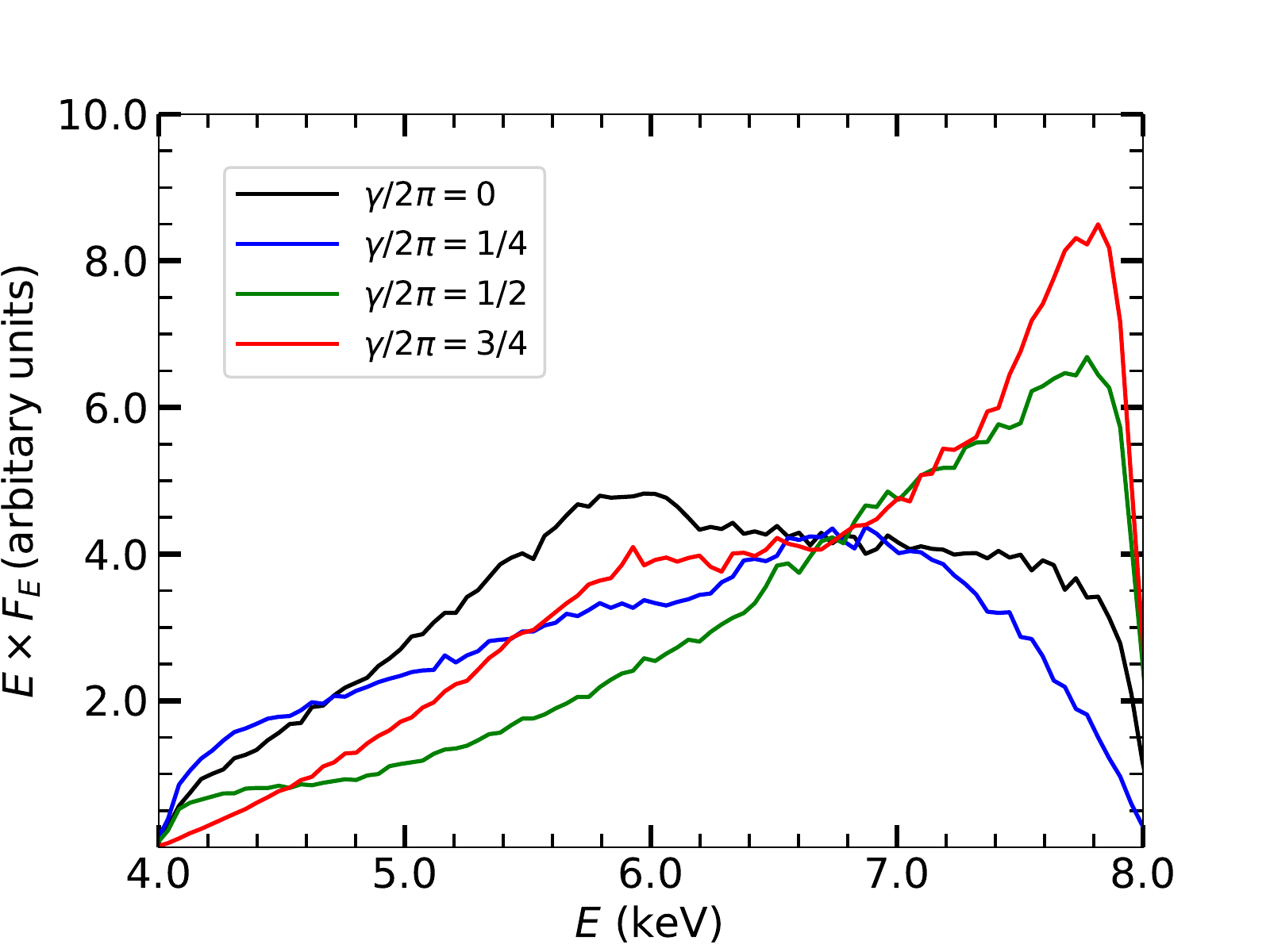}
\caption{
The observed Fe K$\alpha$ line for large truncation radius $R_{\rm tr} = 10$ and middle inclination angle $\cos \theta \sim 0.5$. The black, blue, green, and red profiles correspond to four specific precession angles, $\gamma/2\pi = 0$, $1/4$, $1/2$, and $3/4$, respectively.
The azimuthal angle of the observer $\varphi = \pi$. The BH spin $a=0.3$ is assumed.
\label{line_shape_r10_phi2}}
\end{figure}

We replot the simulation results above but for $\varphi=\pi$.
In Fig. \ref{mid_refl_pattern_r10_phi2}, we plot the reflection patterns taken at four different precession angles $\gamma/2\pi = 0$, $1/4$, $1/2$, and $3/4$ for small truncation radius $R_{\rm tr}$=10, as seen by an observer at the middle viewing angle $\cos \theta \sim 0.5$. The corresponding spectroscopic profiles of the Fe K$\alpha$ line are plotted in Fig. \ref{line_shape_r10_phi2}.
At the first half of the precession period ($\gamma/2\pi <0.5$), the incident luminosity concentrates on the receding side of the disk, from the point of view of the observer. Due to the Keplerian rotation of the disk, however, the flux from the approaching side will be boosted and the flux from the receding side will be suppressed. Therefore, given the combination of these two effects, the reflection pattern shows marignal difference over the disk (see panel (a) and (b)). As a result, the corresponding Fe K$\alpha$ line is roughly flat between 4 and 8 keV (the black and blue lines in Fig. \ref{line_shape_r10_phi2}). However, at the second half of the precession period ($\gamma/2\pi > 0.5$), the incident flux concentrates on the approaching side of the accretion disk where the reflection pattern is much brighter than the receding side (see panel (c) and (d)). Consequently, the blue wing of the emission line is enhanced (red and green lines in Fig. \ref{line_shape_r10_phi2}).

\begin{figure}
\includegraphics[width=\columnwidth]{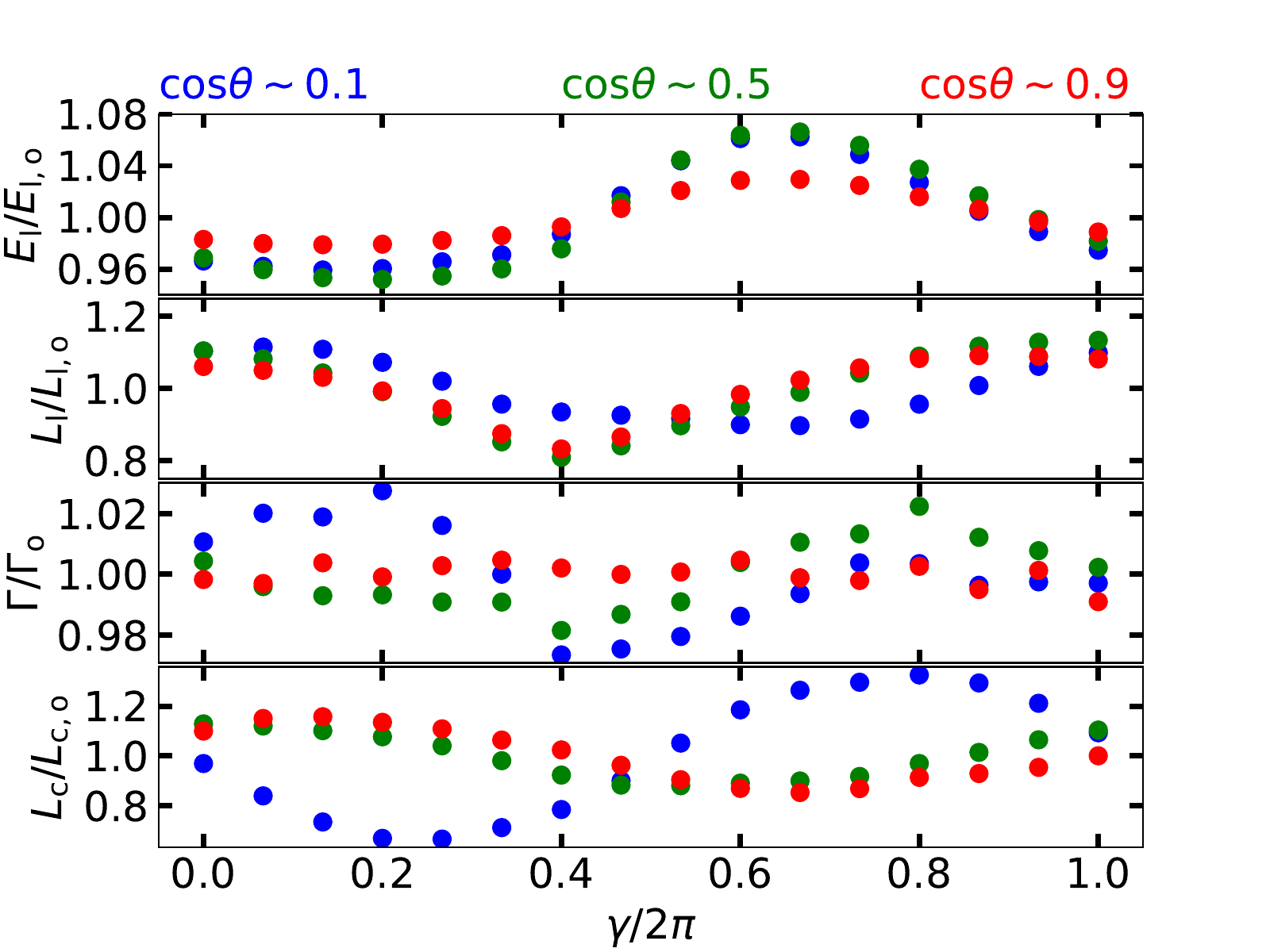} 
\caption{
The spectral properties of the observed 2-10 keV radiation, as a function of precession angle $\gamma/2\pi$, for large truncation radius $R_{\rm tr} = 10$. From top to bottom, each panel corresponds to the central energy of the line $E_{\rm l}$, the iron line flux $L_{\rm l}$, the spectral slope $\Gamma$, and continuum flux $L_{\rm C}$. Correspondingly, the phase-averaged values are labeled as $E_{\rm l,o}$, $L_{\rm l,o}$, $\Gamma_{\rm o}$ and $L_{\rm C,o}$. 
%In order to demonstrate the variation of these spectral parameters, the y-axis in all panels represent the ratio of the phase-dependent value and the phase-averaged value, i.e., $E_{\rm l}/E_{\rm l,o}$, $L_{\rm l}/L_{\rm l,o}$, $\Gamma/\Gamma_{\rm o}$ and $L_{\rm c}/L_{\rm c,o}$.
The blue, green, and red points are for the viewing angles $\cos \theta \sim 0.1$, 0.5, and 0.9, respectively. The azimuthal angle of the observer $\varphi = \pi$. The BH spin $a=0.3$ is assumed.
\label{line_para_r10_phi_2}}
\end{figure}

The spectral analysis between 2 and 10 keV including the Fe K$\alpha$ line is replotted in Fig. \ref{line_para_r10_phi_2}, but for the observer azimuth $\varphi=\pi$. In this case, the variation of the emission-line centroid energy $E_{\rm l}$ can be larger in comparison to the case of $\varphi=0$, up to $\sim$ 5\%. More importantly, when the source is viewed at large inclination angle, in the case of $\varphi$=0 the phase difference between the centroid energy $E_c$ and the line luminosity $L_{\rm l}$ is small with $\Delta \gamma/2\pi < 0.1$; in the case of $\varphi=\pi$, the phase difference between the centroid energy $E_c$ and the line luminosity $L_{\rm l}$ is large with $\Delta \gamma/2\pi > 0.3$. Moreover, for the latter case, the phase difference between the continuum luminosity $L_{\rm l}$ and the line luminosity $L_{\rm l}$ is large with $\Delta \gamma/2\pi \sim 0.4$. Therefore, given this effect, measuring the phase difference between the line flux $F_{\rm line}$ and $E_c$/$F_{\rm 2-10}$ could be a good diagnostic for the azimuthal position of the observer.

\subsection{Effect of the optical depth}
%{\bf It is shown in Fig. ??? that the observed direct flux for the truncation radius $R_{\rm in} = 90$ depends on the inclination angle, since the corona is optically thick with $\tau \sim 2.$ in this case. In order to study the effect of the optical depth on the variability properties, we run the simulation for $R_{\rm in} = 90$ again, but assuming the corona is optically thin with the $\tau \sim 0.5$. In the case of the optical depth, the phase-resolved direct flux turns out to be fairly independent of the inclination. And the variation pattern of the incident flux is identical to the one for the case pf the optically thick, except the overall amplitude of the incident flux in the case of the optically thin is enhenced.} 

In this work, the corona is assumed to be torus-like with $H/R \sim \tan 15^{\circ} \simeq 0.3$. For the truncation radius $R_{\rm tr} = 90$, the corona is optically thick with the maximum optical depth $\tau \simeq 2.1$ (see Sect. 2.2) when the line of sight is perpendicular to the corona axis. As the viewing angle decrease, the effective optical depth (i.e., being integrated over all lines of sight leading to different points within the corona) along the line of sight will then decrease. Therefore, in Fig. \ref{refl_frac}, the observed direct flux for the truncation radius $R_{\rm tr} = 90$ depends on the inclination angle. In contrast, for the truncation radius $R_{\rm tr} = 10$, the corona is optically thin with maximum optical depth $\tau \simeq 1.0$, and then the phase-averaged direct flux is fairly independent of the inclination angle.

\begin{figure*}
\includegraphics[width=2\columnwidth]{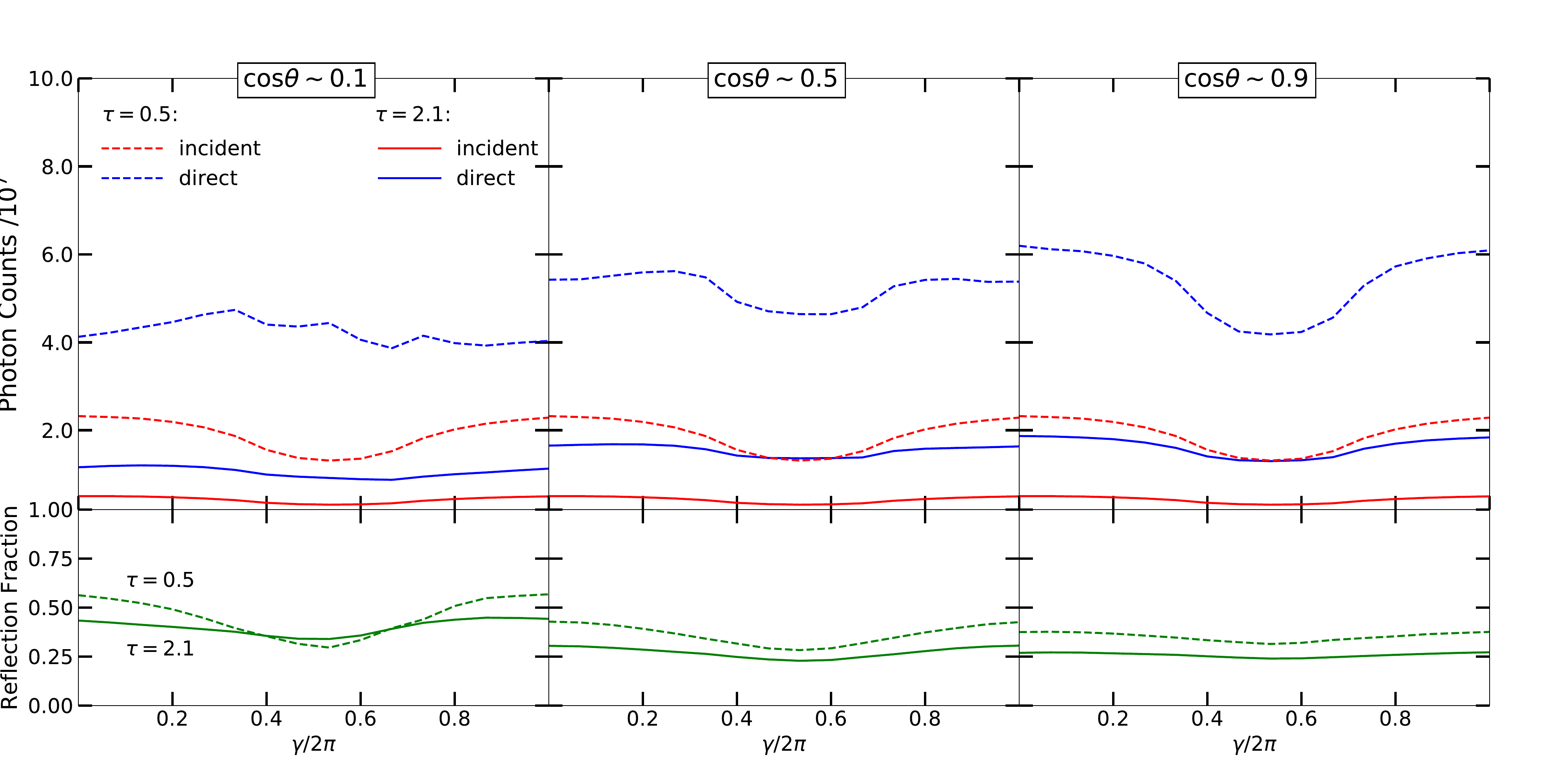} 
\caption{
Upper panels: the normalized photon counts of illuminating the disk (incident; in red color) and being observed by the observer at infinity (direct; in blue color), for $R_{\rm tr}=90$; Bottom panels: the reflection fraction, i.e., the counts ratio of incident photons and direct photons. The solid and dashed lines are for $\tau \simeq 2.1$, and $\tau \simeq 0.5$, respectively. From the left to right panels, the inclination angle of the observer corresponds to $\cos \theta \sim 0.1$, 0.5, and 0.9, respectively. The BH spin $a=0.3$ is assumed.
\label{refl_frac_thin}}
\end{figure*}

In order to study the effect of the optical depth on the variability properties, we run the simulation for $R_{\rm tr} = 90$ again, but assuming the maximum optical depth $\tau \sim 0.5$. In Fig. \ref{refl_frac_thin}, we plot the direct/incident flux and the resultant reflection fraction, as a function of precession phase, for the cases of the optically thin and thick cases. The difference in the phase-averaged direct flux between high and low inclinations turns out to be reduced.
However, the variation pattern of the incident flux is still identical to the one in the case of optically thick corona, and there is no significant change in the patterns of the reflection fraction. Moreover, we check that there is no qualitative difference in the corresponding variation patterns of the disk reflection including Fe K$\alpha$, due to the changes in the optical depth.

\subsection{Comparison to previous work}
The phase-dependent of QPO properties including Fe K$\alpha$, which are simulated in this work, qualitatively agree with the far simpler earlier treatment of Ingram \& Done (2012, hereafter ID12). ID12 also found the behavior displayed in Fig. \ref{line_shape_r90} and \ref{line_shape_r10}, i.e., the line varies in such a way that the red wing can dominate over the blue wing for a large $R_{\rm tr}$, but not for a small $R_{\rm tr}$. There are a few important differences though. ID12 concluded that the bluest line always happened during the rising phase of the continuum flux, and the reddest line always happened during the falling phase of the continuum flux. This is no longer the case with the new analysis for a number of reasons. First, in the new analysis, the bottom of the corona provides the strongest illumination, meaning that the brightest patch of the disk is always in the opposite direction to where the projection of the corona axis on the disk plane points. This means that for the viewer azimuth whereby the corona maximally faces the observer at maximum misalignment (i.e. observer at $\varphi=0$), the bluest line is the falling phase and the reddest line is the rising phase (i.e. the opposite of ID12). Second, we include the effect of variation of seed photons as the misalignment angle between disk and corona varies and also light bending, which was ignored by ID12. This leads to the precession phase of the peak continuum flux depending not only on the angle between the corona axis and the line of sight (as in ID12), but also on the angle between the corona and disk axes. We also include light bending here, which is far more important for $R_{\rm tr} = 10$ than for $R_{\rm tr} = 90$.

\subsection{Comparison to observation}
Ingram et al. (2017, hereafter ID17) applied a phase-resolved analysis to XMM-Newton and NuSTAR observations of H1743-322 to study its spectral properties, including the Comptonization continuum flux, the centroid energy of Fe K$\alpha$ line, and the reflection fraction as a function of QPO phase. It was found that the continuum normalization by the spectral fits varies in the wave-like form with high significance. In this work, we show that the simulated Comptonization flux is modulated due to Keplerian rotation of the corona, with the maximum and minimum fluxes roughly occurring at the first half, $\gamma/2\pi \sim 0.3$, and the second half, $\gamma/2\pi \sim 0.7$ (see Fig. \ref{refl_frac} and the discussion in Sect. 3.1.2), which qualitatively agrees with the results of ID17. 

In ID17, the centroid energy of the Fe K$\alpha$ line of H1743-322 shows a characteristic variation with QPO phase, with two maxima at $\sim 0.2$ and $\sim 0.7$ of the QPO cycles.
However, in this work, we only see one maximum of the iron line energy within one precession cycle. Given that the iron line intrinsically arises from the illumination of the disk in the reflection model, the two maxima of the iron line centroid energy require two bright patches on the disk surface. But we are still only seeing one bright patch in the current simulation (e.g., see Fig. \ref{irradiation_pattern_r90}). This is because, in our current simulation, we assume the corona to be torus-like radially extended. We note that in the case of lamp-post (point-like) geometry, two bright patches on the disk are seen (see Fig. \ref{lamp_post}). If this point-like corona is vertically extended, e.g., a jet-like base (Wilkins et al. 2015; Kara et al. 2019; You et al. 2020, in preparation) and precesses around the BH spin axis with some
period (Liska et al. 2018), then we would expect these two bright patches on the disk to vary, which may explain the two maxima of the iron line in H1743-322.
The simulation for such a geometry of the corona, i.e., a jet base coupling with an extended corona, will be done in detail in our next paper.    

The reflection fraction of H1743-322 in ID17 also varies with QPO phase. We note that H1743-322 was in the hard state according to XMM-Newton and NuSTAR observations in ID17. In our simulation results, we found that the predicted reflection fraction indeed is highly modulated as well, when the truncation radius $R_{\rm tr}$ is small and the inclination to the source is large. However, the modulation of the measured reflection fraction in ID17 is much more complex, in comparison to our results. Given that the reflection fraction intrinsically depends on the relative geometry of the X-ray source with respect to the disk, the complex variation pattern of the reflection fraction indicates the geometry of the X-ray source should be more advanced, as we discuss above, e.g., a jet base coupling with an extended corona.

\section{Summary}
In this work, we studied the phase-resolved irradiation/reflection (including the Fe K$\alpha$ line) and the resultant spectral properties if the X-ray primary source undergoes Lense-Thirring precession around the spin axis of the BH. 
In the truncation geometry of the primary X-ray source, the main conclusions are summarized as follows:
\begin{enumerate}
\item The incident pattern over the outer disk rotates with the precession of the inner corona. Because the direct radiation from the corona varies as a function of the precession phase, the derived reflection fraction will correspondingly vary with the precession phase, which depends on the truncation radius (i.e., the spectral state) and the inclination angle.
\item The reflection off the disk also rotates with the precession, the observed pattern of which depends on the truncation radius and the inclination angle.
\item The Fe K$\alpha$ line profile will correspondingly change with the precession phase. More specifically, the line luminosity and the flux-weighted centroid energy vary with the precession phase. 
\item The continuum luminosity from the precessing corona could apparently lag the line luminosity in phase, if the truncation radius is small when the Doppler effect of the observed radiation becomes significant.
\end{enumerate}  
  
However, the geometry of the primary X-ray source is still uncertain, which may alternatively be described by the lamp-post geometry or as a vertically extended structure, like a jet base. Then, two bright patches are formed on the disk, the flux of which will be modulated by the precession of the X-ray primary source, but likely with half of the precession period. This effect may be responsible for the observed harmonic in the power spectrum of BHXRBs, which
will be further studied in the next paper.

\section{Acknowledgments} 
B.Y. thanks Phil Uttley for discussions and hospitality at the Anton Pannekoek Institute for Astronomy, University of Amsterdam. B.Y. thanks Zhu Liu for discussing the simulation of the Fe K$\alpha$ line. This work is supported by the NSFC (11903024, U1838103, U1931203). This work was partly supported by Polish National Science Center grant 2015/17/B/ST9/03422. A.I. is supported by a Royal Society University Research fellowship. 

{}

\end{document}